\documentclass[reprint,asmsmath,amssymb,aps,prb]{revtex4-2}

\usepackage{bm}

\usepackage{balance}
\usepackage{float}

\usepackage{etoolbox}
\usepackage{threeparttable}

\makeatletter
\patchcmd{\frontmatter@abstract@produce}
  {\vskip200\p@\@plus1fil
   \penalty-200\relax
   \vskip-200\p@\@plus-1fil}
  {}
  {}
  {}
\makeatother

\usepackage{hyperref}

\usepackage{amsmath}
\usepackage{physics}
\usepackage{siunitx}
\usepackage{enumitem}

\usepackage{tikz}
\usepackage{pgfplots}
\pgfplotsset{compat=1.14}
\usetikzlibrary{trees}
\usetikzlibrary{decorations.pathmorphing}
\usetikzlibrary{decorations.markings}
\usetikzlibrary{patterns}

\definecolor{fluor}{RGB}{249, 65, 68}
\definecolor{ee}{RGB}{243, 114, 44}
\definecolor{extra}{RGB}{248, 150, 30}
\definecolor{tbr}{RGB}{249, 199, 79}
\definecolor{eii}{RGB}{144, 190, 109}
\definecolor{aug}{RGB}{67, 170, 139}
\definecolor{pht}{RGB}{87, 117, 144}

\DeclareUnicodeCharacter{2212}{\ensuremath{-}}

\usepackage{amssymb}

\usepackage{array}
\usepackage{arydshln}
\usepackage{booktabs}
\usepackage{tabularx}

\makeatletter
\newcommand*{\smallrel}[2][.8]{%
  \mathrel{\mathpalette{\smallrel@{#1}}{#2}}%
}
\newcommand*{\smallrel@}[3]{%
  \sbox0{$#2\vcenter{}$}%
  \dimen@=\ht0 %
  \raise\dimen@\hbox{%
    \scalebox{#1}{%
      \raise-\dimen@\hbox{$#2#3\m@th$}%
    }%
  }%
}
\makeatother

\begin{document}

\title{Heavy-element damage seeding in proteins under XFEL illumination}

\author{Spencer K. Passmore}
\email{spencerpassmore@swin.edu.au}
\affiliation{School of Physics, University of Melbourne, Parkville, Victoria 3010, Australia}
\affiliation{Optical Sciences Centre, Swinburne University of Technology, Hawthorn, Victoria 3122, Australia}
\author{Alaric L. Sanders}
\affiliation{T.C.M. Group, Cavendish Laboratory, University of Cambridge, Cambridge CB3 0HE, United Kingdom}
\affiliation{School of Physics, University of Melbourne, Parkville, Victoria 3010, Australia}
\author{Andrew V. Martin}
\affiliation{School of Science, STEM College, RMIT University, Melbourne, Victoria 3000, Australia}
\author{Harry M. Quiney}
\affiliation{School of Physics, University of Melbourne, Parkville, Victoria 3010, Australia}

\begin{abstract}

Serial femtosecond X-ray crystallography (SFX) captures the structure and dynamics of biological macromolecules at high spatial and temporal resolutions. The ultrashort pulse produced by an X-ray free electron laser (XFEL) `outruns' much of the radiation damage that impairs conventional crystallography. However, the rapid onset of `electronic damage' due to ionization limits this benefit.
Here, we distinguish the influence of different atomic species on the ionization of protein crystals by employing a plasma code that tracks the unbound electrons as a continuous energy distribution. 
The simulations show that trace quantities of heavy atoms ($Z>10$) contribute a substantial proportion of global radiation damage by rapidly seeding electron ionization cascades. In a typical protein crystal, sulfur atoms and solvated salts induce a substantial fraction of light-atom ionization. In further modeling of various targets, global ionization peaks at photon energies roughly 2 keV above inner-shell absorption edges, as sub-2 keV electrons initiate ionization cascades that are briefer than the XFEL pulse. These results indicate that relatively small quantities of heavy elements can substantially affect global radiation damage in XFEL experiments.
\end{abstract}

\maketitle

\section{Introduction}

Serial femtosecond crystallography (SFX) is a potentially revolutionary structural determination tool for biology that overcomes key challenges faced by conventional X-ray crystallography~\cite{botha&frommeReviewSFX2023}.
Using X-ray free electron lasers (XFELs) to serially illuminate small crystals with ultrabright, femtosecond pulses of radiation, the structural signal of a molecule is captured so swiftly that the atomic nuclei are effectively frozen in place \cite{neutze,chapman2014,suga2015}. 
This approach not only mitigates structural damage, allowing for the use of much higher fluences, but also facilitates time-resolved crystallography. As such, SFX can be applied to capture molecular movies of rapid biological processes, including photoactivated dynamics \cite{tenboerPhotoactiveYellowProtein2014,pandeyTimeresolvedSerialFemtosecond2020,sugaTimeresolvedStudiesMetalloproteins2020,pande,standfussMembraneProteinDynamics2019,carrilloPhytochromePhotocycle2021,suga2017} and enzyme catalysis \cite{dasguptaMixandinjectXFELCrystallography2019,olmos2018,hullCarbonicAnhydraseNMRVsXFELStructures2024}.

In the ideal limit of SFX, the X-ray diffraction pattern of each crystal is captured by an ultrashort pulse before any secondary damage processes can propagate~\cite{spenceOutrunningDamageElectrons2017, seibertSingleMimivirusParticles2011, boutetHighResolutionProteinStructure2012,chapmanStructureDeterminationUsing2017,nass2019}. 
In this `diffraction before destruction' picture, damage no longer constrains the X-ray intensities that allow for successful structure determination.
This freedom allows for structure determination from micron or sub-micron crystals \cite{chapmanFemtosecondXrayProtein2011, chapman2014}. 
Indeed, several XFEL experiments have captured the structures of radiation-sensitive metalloproteins without evidence of significant structural deformation~\cite{chapmanFemtosecondXrayProtein2011,suga2015,sugaTimeresolvedStudiesMetalloproteins2020}. 
The technique can thus be applied to molecules that do not readily form large crystals, such as membrane proteins~\cite{GisrielMembraneProteinEXFEL2019,hunterTowardMembraneProteinDetermination2011,martinj,zhao,botha&frommeReviewSFX2023}. 
Using smaller crystals is particularly advantageous for time-resolved experiments, which demand larger quantities of sample than static structure determination~\cite{standfussMembraneProteinDynamics2019}, are less tolerant to beam attenuation, and benefit from high crystal surface area to volume ratios when attempting to induce structural changes chemically \cite{schmidtMixAndInjectReactionInitiationDiffusion2013}.

In practice, ultrafast electronic processes disrupt the simple diffract-and-destroy picture -- scattering signals seen in experiments generally remain dose-dependent even at the practical minimum pulse widths achievable ($\sim$\;$10~\si{\femto\second}$) ~\cite{martinCoherenceLossSample2016,nassStructuralDynamicsProteins2020,boutetHighResolutionProteinStructure2012,abbeyXrayLaserinducedElectron2016}. In particular, the electrons ejected through photoionization and subsequent Auger decays initiate electron impact ionization (EII) cascades that will go on to become the main cause of ionization well before the end of the pulse~\cite{vinkoCollisionalIonizationImportance2011, chapman2014,caleman2011}.
This `electronic damage' has complex, non-linear effects on the scattered wavefield~\cite{martinCoherenceLossSample2016,quineyBiomolecularImagingElectronic2011,kozlov2020_coherence,abdullahTheoreticalLimitationsXray2018}, and various theoretical treatments have been proposed to model the underlying ultrafast damage processes~\cite{abdullahTheoreticalLimitationsXray2018,quineyBiomolecularImagingElectronic2011,kozlov2020_model,kozlov2020_coherence,hau-riegeNonequilibriumElectronDynamics2013,caleman2011}. Efficient use of limited XFEL beamtime~\cite{GisrielMembraneProteinEXFEL2019,moraSerialSynchrotronRadiationDamageDose2020} therefore often depends on identifying experimental parameters that minimize electronic damage~\cite{dickersonRADDOSE-XFEL2020}.

Heavy atoms ($Z>10$ in this context)  are ubiquitous in protein crystals. Though they typically comprise less than 1\% of the illuminated atoms, they play a key role in the biochemistry of metalloproteins. Additionally, their unusual scattering behavior has various applications to X-ray crystallography. Anomalous phasing, a notable example, has been successfully extended to SFX with microcrystals using sulfur native to proteins \cite{nakane2015,barends2013AnomalousSulfurSignal,nass2021}, or with much heavier atoms deliberately added for a stronger signal \cite{barends2014, hunter2016, gorel2017TwoColour}. 
The rapid ionization of heavy atoms under high fluences induces a strong time dependence in their atomic scattering factors.
This behavior is central to proposed methods of high-intensity radiation-damage-induced phasing (HI-RIP) \cite{son2011_MAD, sonDeterminationMultiwavelengthAnomalous2013, galliReproducibleRadationDamageProcessesIntenseX-Ray2015, galliTowardsRIP,galli,galliSulfurExperiment,nassStructuralDynamicsProteins2020,caleman2020,chapman2014}.

A large part of the existing body of work dedicated to simulating atomic dynamics under XFEL illumination follows either the Monte Carlo molecular dynamics (MD) paradigm~\cite{abbeyXrayLaserinducedElectron2016, jurekXMDYNXATOMVersatile2016, hoLargescaleAtomisticCalculations2017,hoRoleTransientResonances2020} or a plasma approach~\cite{calemanSimulationsBiomolecularNanocrystals2011, leonovTimeDependenceXray2014, kozlov2020_model,ziajaBoltzmannDiamondWDM2023}. These frameworks are in some senses complementary; the former provides a detailed picture of damage on the local scale, while the latter focuses on global statistics~\cite{caleman2020}. In the MD case, individual ions and electrons are treated as classical particles and tracked through space. Such modeling is too computationally demanding for system sizes over 100--1000 atoms, so cannot simulate typically sized proteins in full~\cite{nassStructuralDynamicsProteins2020,abdullah_2016,abdullahTheoreticalLimitationsXray2018,hau-riegeNonequilibriumElectronDynamics2013}. 
In contrast, models that incorporate a zero-dimensional plasma code may capably simulate much larger (or arbitrary) systems~\cite{calemanSimulationsBiomolecularNanocrystals2011, kozlov2020_model, dawodMolDStruct2024, caleman2020}, mitigating possible difficulties in capturing the effect of trace elements.

Several plasma codes that treat the free electrons as non-thermal (i.e. non-Maxwellian) have been applied in studies of single-element, solid-density targets~\cite{royleKineticModelingXray2017,hau-riegeNonequilibriumElectronDynamics2013,leonovTimeDependenceXray2014,leNonThermalCRETIN2019, ziajaBoltzmannDiamondWDM2023, Ren2023NonLTE-CCFLY,shiCCFLY2024}. However, plasma codes used in prior studies to simulate the damage sustained by biomolecules all approximate the free-electron energy distribution with a Maxwellian, under an assumption of instantaneous thermalization. This treatment may miss important effects -- existing work on single-element, solid-density targets quotes thermalisation times in picoseconds~\cite{kitamuraThermalizationDynamicsPrimary2019,khoRelaxationSystemCharged1985,royleKineticModelingXray2017,abdullah_MD_non-equilibrium2018,hau-riegeNonequilibriumElectronDynamics2013,Ren2023NonLTE-CCFLY}.

In this work, we employ a non-thermal plasma physics code to examine electronic damage in protein crystals containing heavy elements without assuming instantaneous thermalization. We combine a custom-built frozen-shell Hartree-Fock code~\cite{kozlovComparisonHartreeFock2019} with a non-standard B-spline approach to solving the Boltzmann equation for the time-dependent non-equilibrium energy distribution of the free electrons. 
The details of this framework are given in Sec.~\ref{sec:ac4dc}. We apply the model to compare how light (C, N, O) and heavy atoms influence the ionization dynamics of biological matter in Sec.~\ref{sec:lysozymeGd}.
We go on to identify how trace quantities of heavy elements in biomolecular targets give rise to species-dependent high-damage regions of the pulse parameter space in Sec.~\ref{sec:deep_absorption_edges}.
We discuss new avenues for mitigating radiation damage implied by these findings in Sec.~\ref{sec:discussion}.

\section{Simulation Method}
\label{sec:ac4dc}

This study introduces a new non-thermal extension of the atomistic collisional-radiative plasma solver \texttt{AC4DC} described in ~\cite{kozlovComparisonHartreeFock2019}. While the previous version assumed free electrons to instantaneously thermalize after collisional-ionization, the new treatment allows for a non-Maxwellian free-electron energy distribution of arbitrary functional form. The scope of this study is restricted to the ionization dynamics, neglecting nuclear motion. This focus is motivated by two points: (I) Ionization drives the ionic motion (Coulomb explosion) during and immediately following the XFEL pulse. (II) Studies of XFEL radiation damage consistently show changes in ionic state (electronic damage) make up the bulk of the radiation damage for pulses on the order of 10 fs~\cite{abdullahTheoreticalLimitationsXray2018,neutze,nassStructuralDynamicsProteins2020}.

\begin{figure}
	\centering
	\tikzset{
    photon/.style={very thick, decorate, decoration={snake, segment length=7mm,
		amplitude=3mm}},
    electron/.style={very thick, postaction={decorate},
				decoration={markings,mark=at position .55 with {\arrow[scale=1.5]{>}}}},
    gluon/.style={decorate, draw=magenta,
        decoration={coil,amplitude=4pt, segment length=5pt}}
}

\begin{tikzpicture}[scale=0.4]
    \pgfmathsetmacro{\circr}{0.3}

		\draw (0,0) node [anchor=east] {$n=\infty$};
		\draw (0,-4.025) node [anchor=east] {$n=1$};
		\draw (0,-1.05) node [anchor=east] {$n=2$};
		\draw[pattern=dots] (0,0) rectangle ++(16,-0.5);
		\draw[pattern=crosshatch] (0,-0.95) rectangle ++(16,-0.2);
		\filldraw (0,-4) rectangle ++(16,-0.05);
		\draw[very thick, ->] (0,-5) -- (0,6.5) node [anchor=north east]{$E$};
		\draw[pht] (2,-4) circle (\circr);
		\draw[photon, pht] (0.5,2) -- (2,-4);
		\draw[pht] (2.5,-4.1) node [anchor=north,xshift=5]{Photoionisation};
		\draw[electron, pht] (2,-4) -- (3, 6);
		\draw[aug, thick] (5,-4) circle (\circr);
		\filldraw[aug] (5,-1) circle(\circr);
        \draw[electron, aug] (5,-1) -- (5,-4);
		\filldraw[aug] (4,-1) circle(\circr);
        \draw[electron,aug] (4,-1) -- (5,3);
		\draw[aug] (5.2, -3) node[anchor=west]{Auger decay};
		\draw[electron, eii, fill=eii] (6,3.5) -- (7,-1) circle (\circr) coordinate(eii);
		\draw[eii] (eii)++(0,3.5) node[anchor=south]{EII};
		\draw[eii, electron] (eii) -- (9,1.1);
		\draw[eii, electron] (eii) -- (9,2.5);
        \draw (14, -1) circle (\circr) coordinate(tbr);
        \draw[electron, tbr] (12, 0.5) -- (tbr);
		\draw[tbr] (13.5,2) node[anchor=south]{TBR};
		\draw[electron, tbr] (12, 1.5) -- (14, -1);
		\draw[electron, tbr] (14, -1) -- (15.5, 2.5);
		\filldraw[fluor] (10,-1) coordinate(F1) circle (\circr);
        \draw[electron, fluor] (F1)  -- (10,-4);
		\draw[fluor, very thick] (10,-4) circle(\circr);
		\draw[fluor] (11,-2.5) node [anchor=south west]{Fluorescence};
		\draw[photon, fluor] (10,-4) -- (16,-3);
		\draw[electron, ee] (6.6,6) -- (10,5);
		\draw[electron, ee] (7,4.5) -- (10,5);
		\draw[electron, ee] (10,5) -- (15,5.5);
		\draw[electron, ee] (10,5) -- (16,4.5);
		\draw[ee] (11,6) node {EE};

		\draw[decorate,decoration={brace,amplitude=10pt},yshift=0pt](16,-0.5)
		-- (16,-4.1) node [black,midway, anchor=west, xshift=10pt]{$P_\xi(t)$};
		\draw[decorate,decoration={brace,amplitude=10pt},yshift=0pt](16,6)
		-- (16,0) node [black,midway, anchor=west, xshift=10pt]{$f(\epsilon, t)$};
	\end{tikzpicture}
	\caption{Bound-free, bound-bound, and free-free transitions in biomolecular plasma. Acronyms denote electron-electron scattering (EE) electron-impact ionization (EII) and three-body recombination (TBR). Bound energy levels are reminiscent of carbon for illustrative purposes. Processes are labeled to indicate the populations they affect: one or both of $\bm{P}$ (the ionic states) and $f$ (the free electrons). The dotted section at $E=0$ represents the weakly-bound molecular structure that is ignored here. Filled circles represent initial-state bound electrons, and hollow circles their final states.}\label{fig:transitions}
\end{figure}

The model couples the free-electron energy distribution, $f(\epsilon, t)$, to the population of possible ionic states, $P_\xi(t)$, through the processes of photoionization, Auger and fluorescent decay, electron impact ionization (EII), three-body recombination (TBR), and pairwise Coulomb electron-electron interactions (EE).
These processes are summarized in Fig.~\ref{fig:transitions}. Atomic parameters are calculated in the radially-averaged Hartree-Fock approximation~\cite{kozlov2020_model,kozlovComparisonHartreeFock2019}. EII and TBR are approximated using the well-established binary-encounter dipole model of Kim and Rudd~\cite{kimBinaryencounterdipoleModelElectronimpact1994}. The equations of motion, obtained from radially averaging the Boltzmann equation, then read

\begin{align}
\frac{\partial}{\partial t} f(\epsilon, t)
    &= \mathcal{Q}[P_\xi, f](\epsilon)
 \label{eq:Boltzmann}
 \\
 \frac{d}{dt} P_\xi(t) &= \sum_{\eta\neq\xi} \Gamma_{\eta\to\xi} P_\eta(t) - \Gamma_{\xi\to\eta}P_\xi(t)\ ,
 \label{eq:Master}
\end{align}
where $\xi$ and $\eta$ denote electron configurations, and $\mathcal{Q}[P_\xi, f](\epsilon)$ and $\Gamma$ represent the couplings to processes affecting $f$ and $P$, respectively (illustrated in Fig.~\ref{fig:transitions}). Details for the calculations of the atomic cross-sections, $\Gamma$, are given in Ref.~\cite{kozlovComparisonHartreeFock2019}. In terms of separate processes, $\mathcal{Q}[P_\xi, f](\epsilon)$ has the form
\begin{align}
    \mathcal{Q}[P_\xi, f](\epsilon) =& 
    \mathcal{Q}^{\rm Photo}[P_\xi](\epsilon) 
    + \mathcal{Q}^{\rm Auger}[P_\xi](\epsilon)
    +\mathcal{Q}^{\rm EII}[P_\xi,f](\epsilon)\nonumber \\
    &+\mathcal{Q}^{\rm TBR}[P_\xi,f,f](\epsilon)
    +\mathcal{Q}^{\rm EE}[f,f](\epsilon)~.
    \label{eq:Qee_explicit}
\end{align}
The notation adopted in Eq.~\eqref{eq:Qee_explicit}
is such that each $\mathcal{Q}[...]$ is multilinear in all arguments. Explicit expressions for these source terms and collision kernels are given in Appendix~\ref{app:numerics}.

The new code tracks an arbitrary non-Maxwellian distribution by expanding $f(\epsilon)$ with piecewise-polynomial B-splines $B_k(\epsilon)$~\cite{boorPracticalGuideSplines1978}.
These basis functions are non-zero only for a small region of energy, allowing for an efficient sparse internal representation of $\mathcal{Q}$ without sacrificing the differentiability of $f$. 
As the simulation progresses, the spline grid is dynamically adapted to have increased density across the energy range spanning the bulk of the thermal electrons and in the vicinity of the main primary ionization peaks (see App.~\ref{app:numerics} for further details).   
This adaptive grid allowed the code to perform full dynamical non-thermal plasma simulations of lysozyme in under an hour on a contemporary desktop. 
 
The total number of modeled atomic configurations for an element scales like the factorial of the total number of subshells of the ground state, making heavier elements very costly to simulate. For computational expedience, $n>=2$ shells of $Z\geq30$ elements were modeled with a single angular momentum (i.e. a single energy level). 
Test simulations showed ionization rates of light and heavy elements were barely affected by this approximation in the regime considered in this work
(see Fig.~\ref{fig:fe_single_shell_comparison}).

We quantitatively compared this new version of \texttt{AC4DC} against a number of published results in the literature, which are presented in App.~\ref{app:comparisons}. 
The code achieved excellent agreement with non-Maxwellian Monte-Carlo simulations by \texttt{ddcMD} of amorphous carbon~\cite{hau-riegeNonequilibriumElectronDynamics2013} for the evolution of the ionic states and free electrons, and predicted similar ionic state dynamics to simulations by \texttt{XMDYN} of a glycine crystal~\cite{jurekXMDYNXATOMVersatile2016,abdullahTheoreticalLimitationsXray2018}. Somewhat surprisingly, the code also sees good agreement with the particle-in-cell DFT code \texttt{PICLS}~\cite{royleKineticModelingXray2017} when modeling aluminum plasma for both the ion and free-electron populations. \texttt{AC4DC} did not reproduce results for simulations of silicon~\cite{leonovTimeDependenceXray2014} by a plasma code that uses a similar physical framework to \texttt{AC4DC}, but which fits $f(\epsilon)$ to a standard grid rather than using adaptive splines (note we found the adaptive grid crucial to achieving convergence, see App.~\ref{app:numerics}).

\onecolumngrid

\begin{figure}[H]
\centering
\includegraphics[width=0.77\columnwidth]{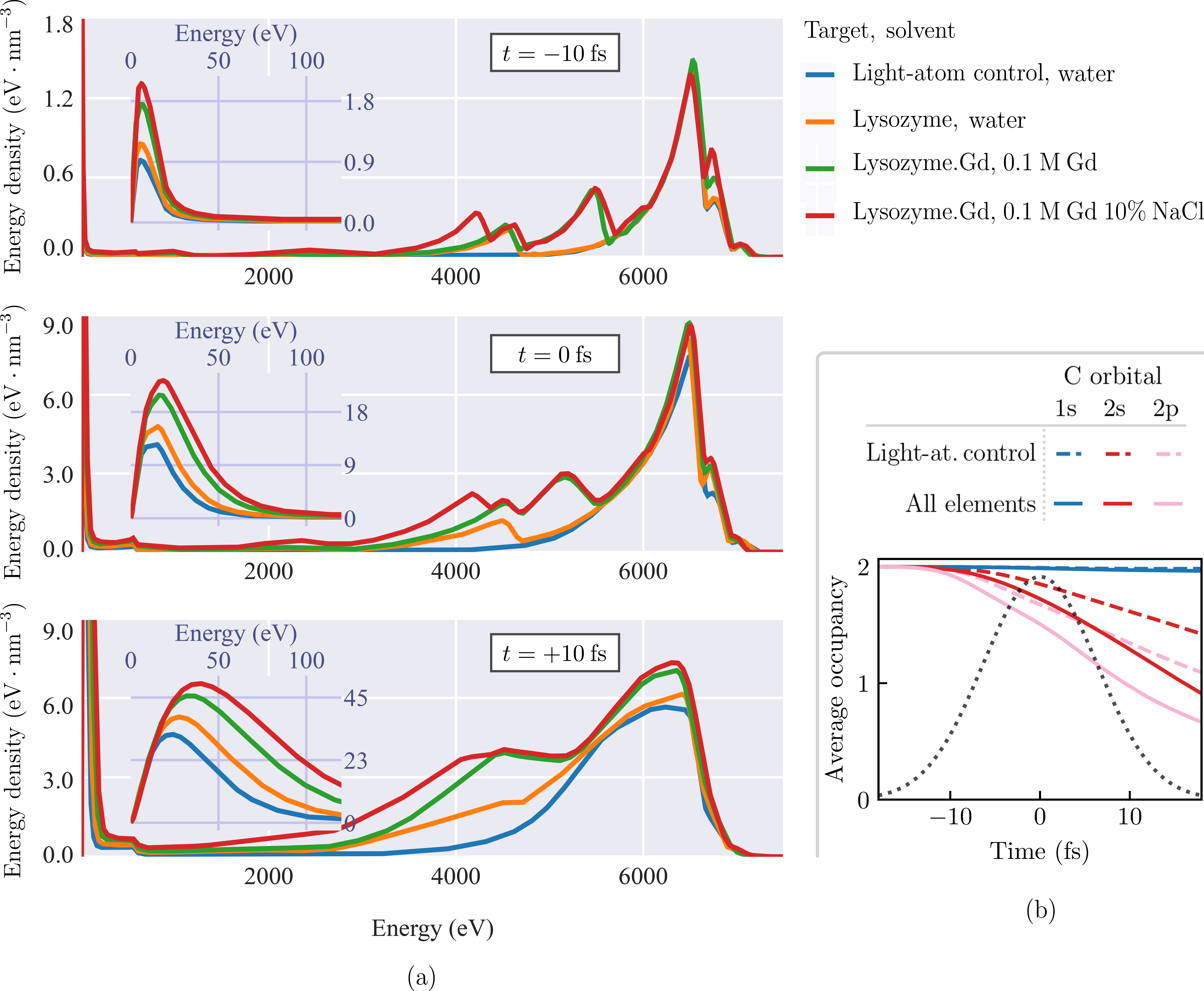}
\caption{
Effect of heavy atoms on electronic damage in lysozyme. (a) Snapshots of the free-electron energy distribution for the light-atom control (blue), lysozyme in water (orange), lysozyme.Gd in 0.1 M Gd solvent (green), and the target of lysozyme.Gd in 0.1 M Gd 10\% NaCl~\cite{nassStructuralDynamicsProteins2020} (red). Inset plots show the distributions at the scale of the thermalized electrons. 
Snapshot times are denoted relative to the pulse's peak intensity ($t=0$ fs). (b) Corresponding evolution of the average occupancy of the electron orbitals of carbon in the light-atom control (broken lines) and lysozyme.Gd (solid lines); the black dotted line traces the temporal pulse profile. Each simulation used a 15 fs FWHM Gaussian pulse with a fluence of $1.75 \times 10^{12}$ 7.112 keV ph$\cdot$\si{\micro\metre}$^{-2}$. }
\label{fig:free_bound_combined} 
\end{figure}

\twocolumngrid

\begin{figure*}
    \makebox[\linewidth][c]{
    \includegraphics[width=7.24409436834in]{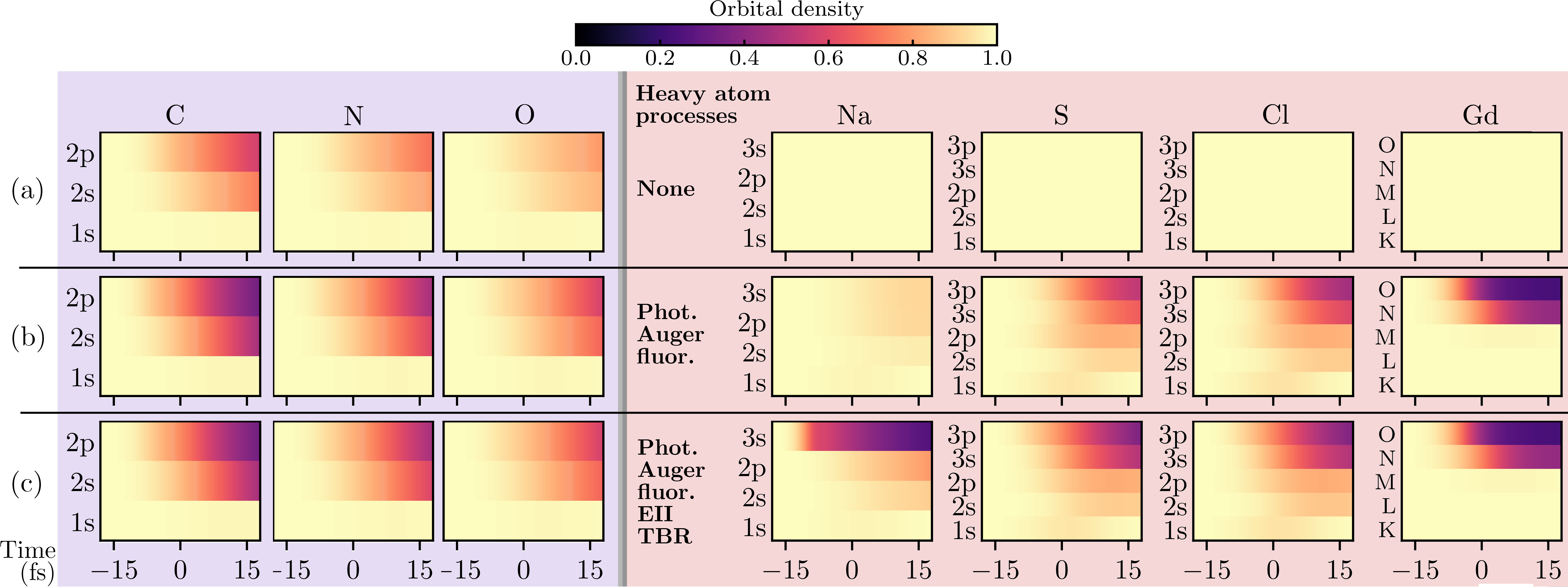}}
	\caption{Impact of heavy-element electronic processes on global ionization for lysozyme.Gd in 0.1 M Gd 10\% NaCl solvent. Plots show average orbital (1s, 2s, \dots) or shell (K, L, \dots) densities normalized to 1 in the initial undamaged state, for each element in the target. All atomic processes in the physical model are enabled for light atoms (C, N, O); while the modeled processes in heavy atoms (S, Gd, Na, Cl) are denoted:
    (a) No heavy atom processes. (b) Heavy atom photoionization, Auger decay, and fluorescence. (c) All atomic processes in the physical model. 
    The elevated depletion of the light-atom orbitals in (b) and (c) is caused by additional collisional-ionization processes stemming from heavy-atom primary ionization. No such difference in the light-atom states is apparent between (b) and (c), indicating heavy-atom secondary ionization is unimportant to light-atom ionization. Pulse parameters match those given in Fig.~\ref{fig:free_bound_combined}.
        }
    \label{fig:orbital_densities}
\end{figure*}

\section{Heavy-seeded ionization cascades -- lysozyme.Gd} %
\label{sec:lysozymeGd}

We first study the impact of heavy atoms in a representative model system -- gadolinium-derivative hen egg-white lysozyme (lysozyme.Gd, derived from PDB entry 4ET8~\cite{pdb_4et8_entry, boutetHighResolutionProteinStructure2012}) -- subjected to a 15 fs FWHM Gaussian pulse of fluence $1.75\times 10^{12}~7.112~\si{keV}$
ph $\cdot$\si{\micro\metre}$^{-2}$. The target contains 35.1\%~(v/v) solvent (10\%~m/v NaCl, 0.1 M Na acetate, 0.1 M Gd~\cite{nassStructuralDynamicsProteins2020}), which we hereafter refer to as the 0.1 M Gd 10\% NaCl solvent for brevity. The resulting unit cell, including disordered solvent atoms, has a chemical composition of \mbox{H$_{13\;259}$C$_{5153}$N$_{1596}$O$_{4009}$S$_{80}$Gd$_{21}$Na$_{93}$Cl$_{87}$}.

 Fig.~\ref{fig:free_bound_combined}(a) shows the free-electron energy distribution of the lysozyme.Gd system at times -10 fs, 0 fs, and +10 fs from the pulse peak. To infer the effect of the heavy atoms on the dynamics, simulations of three `toy' variants of the system are also shown: lysozyme.Gd in 0.1 M Gd solution (H$_{13\;846}$C$_{5143}$N$_{1596}$O$_{4300}$S$_{80}$Gd$_{21}$), lysozyme in water (H$_{13\;942}$C$_{5056}$N$_{1576}$O$_{4386}$S$_{80}$), and a `light-atom control' (H$_{13\;942}$C$_{5056}$N$_{1656}$O$_{4386}$). These have the same solvent density (1.1 g/mol) and protein concentration as the complete system. It can be seen in Fig.~\ref{fig:free_bound_combined}(a) that heavy elements make significant contributions to the high-energy region of the distribution relative to their small presence. In addition, the inset plots show a substantially larger thermal (low-energy) electron population for the complete target than the light-atom control. The vast majority of the additional thermal electrons come from additional secondary ionization events in the \textit{light} atoms. This is reflected by Fig.~\ref{fig:free_bound_combined}(b), which shows that the carbon 2s and 2p orbitals deplete much more rapidly in the complete target.

\subsection{Comparing the ionization behavior of light and heavy elements} 
\label{subsec:ionization_modes}

Fig.~\ref{fig:orbital_densities} shows that the influence of the heavy ions on global ionization almost entirely originates with their primary ionization processes. Modeling that neglects the secondary ionization of heavy atoms (Fig.~\ref{fig:orbital_densities}(b)) still produces a nearly identical evolution in the carbon ionic states to those of the complete simulation shown in Fig.~\ref{fig:orbital_densities}(c). In other words, the effect of heavy elements on the dynamics is almost entirely due to their primary electron emissions.

The primary electron (combined Auger and photoelectron) production rate rises substantially with higher $Z$. 
This is partly due to the huge scaling of the photoabsorption cross-section ($Z^5$ for the K-shell~\cite{Bethe&Salpeter1957}), and partly due to shorter Auger lifetimes of heavier elements ($\lesssim 1$ fs for elements heavier than sulfur) ~\cite{nass2019,campbellAtomicWidths2001,sonFrustratedAbsorption2020}. Fig.~\ref{fig:orbital_densities}(c) shows the M (3n) shells of the Gd ions are nearly full for the entire simulation. The rapid ejection of photoelectrons from the M shell is sustained by the electrons in higher orbitals, which fill core holes on a subfemtosecond timescale through Auger decay and fluorescence. Previous experimental studies have noted such `Auger cycling' significantly elevates primary ionization under XFEL pulses~\cite{hoRoleTransientResonances2020,rudenkoFemtosecondResponsePolyatomic2017}, though here it also elevates the number of secondary ionization cascades seeded by Gd.

To quantify the contribution of each element to global ionization in terms of the damage processes that they seed through primary ionization, we construct a measure of the response of the full free-electron energy distribution $f$ to a particular source of primary electrons.
We introduce auxiliary distributions $g_n(\epsilon, t)$ that partition $f$, i.e. $f(\epsilon, t) = \sum_n g_n(\epsilon, t)$, such that they represent the free-electron density associated with specific primary ionization processes, e.g. $n=$ C Auger, or $n=$ O photoionization. Explicitly, we decompose the right-hand side of Eq.~\eqref{eq:Qee_explicit} as
\begin{align}
    \frac{\partial g_n(\epsilon, t)}{\partial t} &=
    \mathcal{Q}^{\text{Photo}}_n[P]
    +
    \mathcal{Q}^{\text{Auger}}_n[P]
    +
    \mathcal{Q}^{\text{EII}}[P,g_n] \nonumber \\
    &+ 
    \frac{1}{2}\sum_m\Big(
    \mathcal{Q}^{\text{TBR}}[P,g_n,g_m]
   +\mathcal{Q}^{\text{TBR}}[P,g_m,g_n] \nonumber\\
   &+\mathcal{Q}^{\text{EE}}[g_n,g_m]
   +\mathcal{Q}^{\text{EE}}[g_m,g_n] 
    \Big)
    \label{eq:separate_pop}
\end{align}
This choice of decomposition assigns all EII-sourced secondary electrons to the cascade of the impactor ($g_n$), while the density arising from TBR and EE processes between different cascades is evenly divided between the two.

\begin{figure} 
\centering
\includegraphics[]{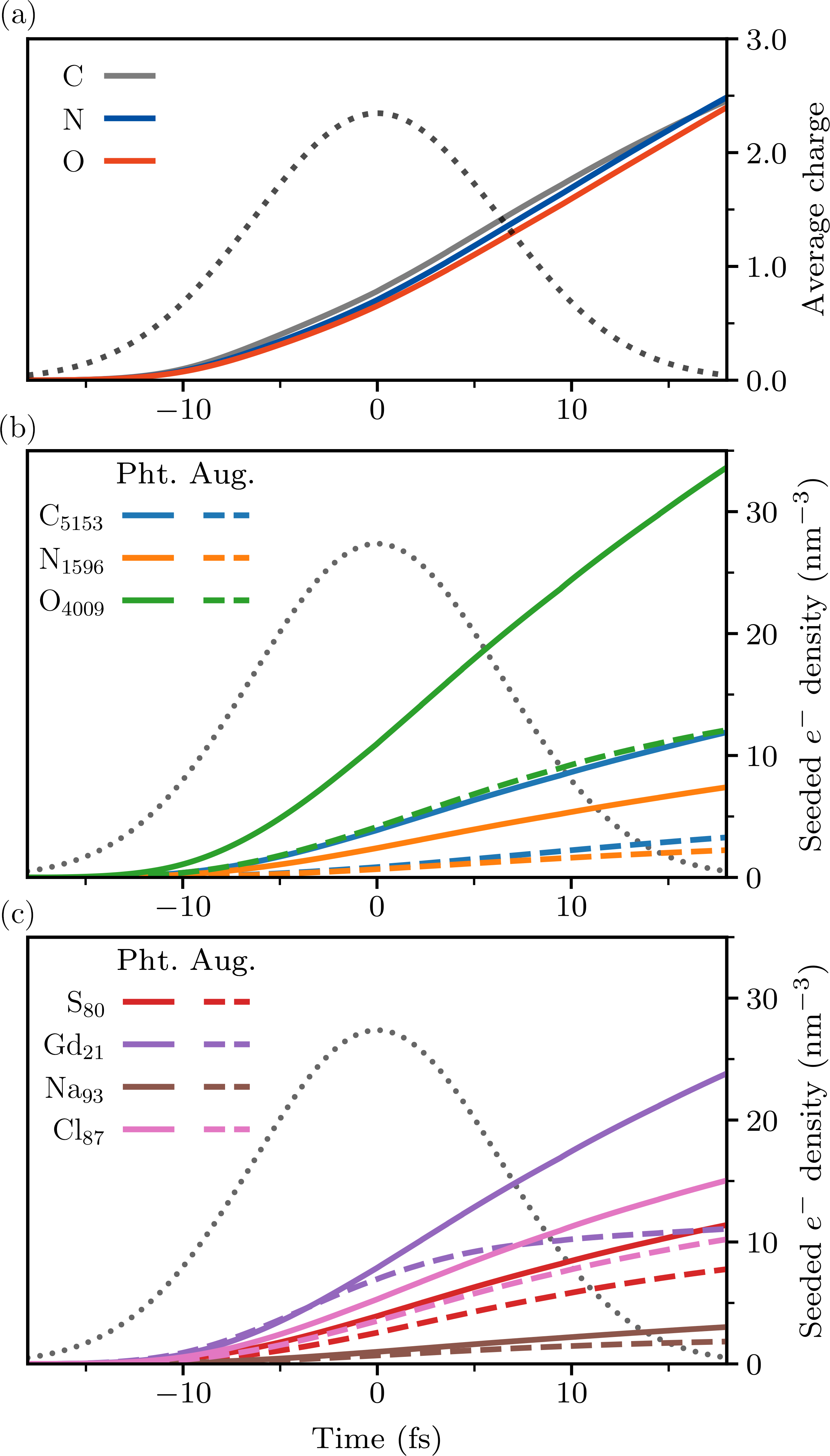}

\caption{
Contribution of each element to the global ionization in the lysozyme.Gd target. (a) plots the average charge of each non-H light-atom species. (b) and (c) trace the collective free-electron density of the EII cascades seeded by each light-atom species and heavy-atom species, respectively. The traces in (b) and (c) correspond to all cascades initiated by photoionization (solid) or Auger decay (dashed) within the population of the denoted atomic species (subscripts in the legend give the population per unit cell). Note these traces sum to the total free-electron density of the system, minus the negligible hydrogen photoelectron density. It can be seen that both photo- and Auger electrons ejected by heavy elements are significant to the global ionization dynamics. At all times, the heavy-element cascades (c) constitute the majority of the free-electron density (58\% at $t=0$). Secondary ionization of heavy elements was ignored in these simulations. Pulse parameters match those given in Fig.~\ref{fig:free_bound_combined}.
} %
\label{fig:cascade_electrons_by_element}
\end{figure}

Fig.~\ref{fig:cascade_electrons_by_element} shows the aggregate density of the cascades seeded by each atomic-species population.
Comparing the light-atom charges (Fig.~\ref{fig:cascade_electrons_by_element}(a)) with the density of the light-atom-seeded cascades (Fig.~\ref{fig:cascade_electrons_by_element}(b))  highlights that the global ionization induced by an atomic species is decoupled from its ionization rate. C, N, and O are ionized at similar rates, largely by EII; however, the O atoms seed a significantly larger fraction of the EII cascades, owing to the $Z^5$ scaling of the K-shell photoabsorption cross-section. 
It can be further seen that the free-electron density seeded by each heavy-element population (Fig.~\ref{fig:cascade_electrons_by_element}(c)) is comparable to that seeded by the far more abundant light-atom populations. Strikingly, the combined density of the Gd-seeded cascades is near or above that of the O-seeded cascades at all times, despite the system containing O and Gd in a ratio of 191:1. Combined, the cascades seeded by heavy elements account for the majority of the freed electrons in the system for the entirety of the pulse.

From this perspective, it can be understood why the heavy elements have a disproportionate influence on the light-atom ionization. The rapid primary ionization of heavy atoms means they seed a significant number of EII cascades in the target despite their trace presence. Since the vast majority of the ionization in the target is due to EII~\cite{vinkoCollisionalIonizationImportance2011,chapman2014}, a substantial fraction of ionization in the light-atom structure occurs through cascades seeded by the heavy elements. %

Note that the density of the cascades seeded by Gd Auger electrons (Fig.~\ref{fig:cascade_electrons_by_element}(b)) sees negligible growth after $t\approx 0$ as a result of the depletion of the O- and N-shells, observable in Fig.~\ref{fig:orbital_densities}(c).
By $t=5$ fs, the average Gd ion charge is above +20. In reality, the highly charged Gd ions would be replenished by bound electrons transported from neighboring atoms~\cite{rudenkoFemtosecondResponsePolyatomic2017}, likely leading to a stronger damage seeding effect.

\subsection{Examining unusual Gd charge measurements in experiments}

\begin{figure}
    \includegraphics[width=3.5in]{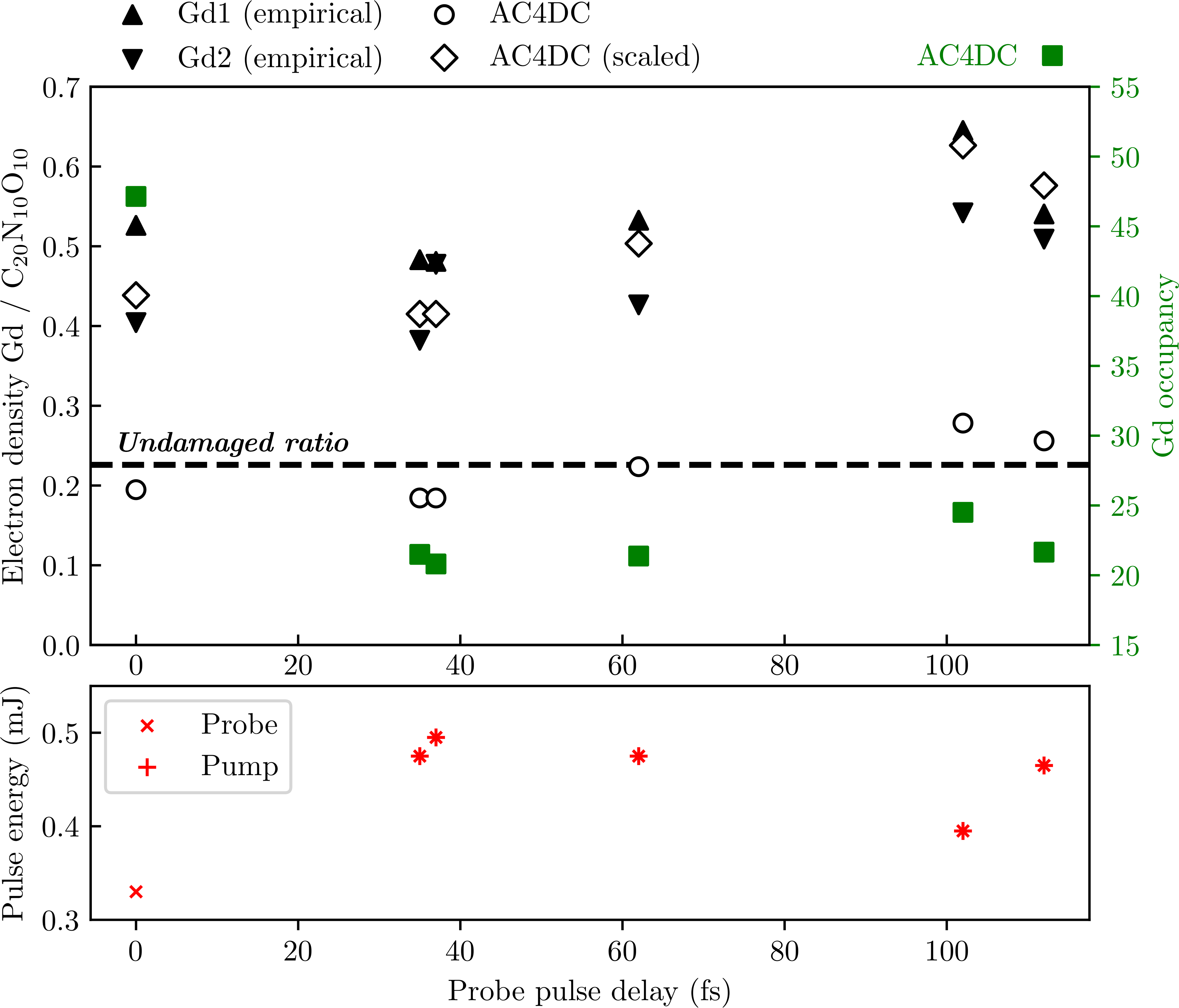}
    \caption{Electron density ratio of Gd ions to light ions (EDR) in lysozyme.Gd as measured in experiments by Ref.~\cite{nassStructuralDynamicsProteins2020} and in corresponding \texttt{AC4DC} simulations. The pulse energy corresponds to individual pump or probe pulses. All data points correspond to a pump pulse at $t=0$ and a probe pulse at some delay, except for the `single pulse' at the $t=0$ data point. Square points correspond to the probe-intensity-averaged Gd occupancy in the simulations (right axis). The dashed line shows the notional EDR of the undamaged target. The triangles correspond to the EDR observed by Ref.~\cite{nassStructuralDynamicsProteins2020} for the two Gd atoms in the asymmetric unit (Gd1 and Gd2). The simulated EDRs (circles) were computed as the probe-intensity-averaged occupancy of a Gd ion over that of a set of light atoms (C$_{20}$N$_{10}$O$_{10}$), as in the considered study~\cite{nassStructuralDynamicsProteins2020}. The diamonds show the simulated EDRs scaled by a constant factor. The relative changes in the empirical EDR between probe pulse delays are remarkably similar to those predicted by \texttt{AC4DC}. In contrast, the EDRs are incommensurate with the Gd occupancy. The fluence of each pulse was modeled as the average within the nominal 0.2~\si{\micro\metre} focus of the experiment, and with the nominal 15 fs FWHM Gaussian temporal profiles. Simulations spanned -18 fs to +18 fs relative to the first and last pulse, respectively. Theoretical values were measured by integrating over a 30 fs timespan, centered on the probe pulse.} %
    \label{fig:occupancy_ratio}

\end{figure}

 The ionization of Gd ions in lysozyme.Gd under XFEL pulses has previously been investigated by Refs.~\cite{nassStructuralDynamicsProteins2020} and~\cite{galli}. The charge density of the Gd ions is inferred from the integrated electron density around the Gd site relative to a reference region of light atoms, which we refer to as the electron density ratio (EDR). Nominally, this is proportional to
\begin{equation}
\label{eq:EDR}
\text{EDR}\propto\int dt \Phi(t)\frac{n_{e}^{Gd}(t)}{2n_{e}^{C}(t)+n_{e}^{N}(t)+n_{e}^{O}(t)},
\end{equation}
where $\Phi(t)$ is the normalized temporal pulse profile, and $n_{e}^{X}(t)$ is the average number of electrons bound to element $X$ in the Gd site or light-atom reference region. Assuming the intensity-averaged light-atom ionization is small, the charge gained by Gd can be approximated. 
 Notably, this measure indicated the ionization of Gd to be unexpectedly low in both studies. Fig.~\ref{fig:occupancy_ratio} compares EDR values obtained from pump-probe experiments by Ref.~\cite{nassStructuralDynamicsProteins2020} (using a light-atom reference region of C$_{20}$N$_{10}$O$_{10}$) with predictions from simulations of these experiments by \texttt{AC4DC} using the solvated lysozyme.Gd target. The simulated EDRs accurately reproduce the features of the experimental data, but rescaled by a factor of $\sim 2$.

\clearpage

\onecolumngrid

\twocolumngrid

 For the four EDR values where the total pulse energy has a range of 0.95--0.99 mJ (35 fs, 37 fs, 62 fs, and 112 fs), the EDR rises with increasing probe delay. %
 We argue this is consistent with the picture of the dynamics described in Sec.~\ref{subsec:ionization_modes}, which highlights that secondary ionization -- the only ionization during the `dead time' between pump and probe -- affects the light atoms far more strongly than the heavy atoms. This implies that the EDR does not gauge the electron density of the heavy atoms but instead serves as a rough indicator of the ratio of secondary ionization to primary ionization. Under this perspective, the EDRs at 0 fs and 102 fs -- which correspond to lower total pulse energies -- are raised due to the greater significance of secondary ionization at lower fluences. 

 It is difficult to diagnose the cause of the factor of 2 systematic offset between the predicted and observed EDRs in Fig.~\ref{fig:occupancy_ratio}. A likely contributing factor is the assumption of a homogeneous spatial profile in the \texttt{AC4DC} model. In the experiment, the beam was nominally focused to a $\sim0.2$~\si{\micro\metre} FWHM Gaussian, smaller than the $\sim0.5$ \si{\micro\metre} lysozyme.Gd nanocrystals~\cite{nassStructuralDynamicsProteins2020}. Photons scattering outside the focus will sample a less-ionized region where, repeating the reasoning from earlier, secondary ionization will be relatively strong and the EDR will be higher. Prior work has highlighted that such photons can make up a substantial fraction of detected photons  ~\cite{murphy2014,galli,nassFerredoxin2015}, suggesting this correction could be quite strong. An additional cause may be bound-to-bound electron transport, which \texttt{AC4DC} neglects. As alluded to in Sec.~\ref{subsec:ionization_modes}, electron transport was previously shown to slow the charge gain of iodine in iodobenzane~\cite{rudenkoFemtosecondResponsePolyatomic2017}, a molecule with fewer light atoms than the Gd complex. Lastly, it is possible that the experimental measurements of the light-atom reference region did not capture all bound electrons. We note the observed EDRs suggest implausibly high ionization rates: The higher EDR observed by the $t=0$ pulse corresponds to an \textit{intensity-averaged} light-atom charge of +3.82, ignoring Gd ionization. For reference, the considered work predicted the average light-atom charge to remain below +1.6 at all times.

Using a similar methodology to Ref.~\cite{nassStructuralDynamicsProteins2020}, Ref.~\cite{galli} also attempted to measure the charge of Gd in two cases -- `high fluence' and `low fluence', with fluences peaking at 7.8 and $0.13\times 10^{12}$ ph$\cdot$\si{\micro\metre}$^{-2}$ respectively. The work reports only the difference in intensity-averaged Gd charge between these cases, termed the Gd `ionization contrast', as measured by computing the difference in the EDRs and assuming negligible light-atom ionization. 
A Gd ionization contrast of 8.8--12 electrons per Gd was observed; however, theoretical modeling using the~\texttt{XATOM} toolkit predicted that the Gd charge difference ought to be 25 under a spatially homogeneous intensity. We repeated this modeling using \texttt{AC4DC}, finding an even more extreme 33.1. This apparent discrepancy between theory and experiment is substantially reduced by accounting for light-atom ionization lowering the denominator in Eq.~\eqref{eq:EDR} (see Fig.~\ref{fig:galli_charge}) -- we obtain a Gd ionization contrast of 21.4 after making this correction. Modeling the cascades seeded by the salt (namely Cl) and Gd ions in the crystal was crucial to seeing this level of agreement, due to their significant contribution to light-atom ionization. Without doing so, the simulated Gd ionization contrast only falls to 25.2 (note the actual Gd charge difference remains at 33.1 in either case).

\section{Damage by different trace elements} 
\label{sec:deep_absorption_edges}

In this section, we present the results of simulations designed to examine how the energy-level structure of a trace, heavy element in a biological target affects electronic damage. Instead of modeling a specific protein, we consider a set of targets with densities of $\sim 1.2$ g cm$^{-3}$ and atomic ratios of C$_{613}$N$_{193}$O$_{185}$X$_{10}$, with the `dopant' X swapped out for various elements. The ionization of hydrogen was ignored for these simulations; testing showed this to be a minor approximation, increasing the average light-atom charge by $\sim$0.1 or less.

\begin{figure}
\centering
\includegraphics[width=7.10694cm]{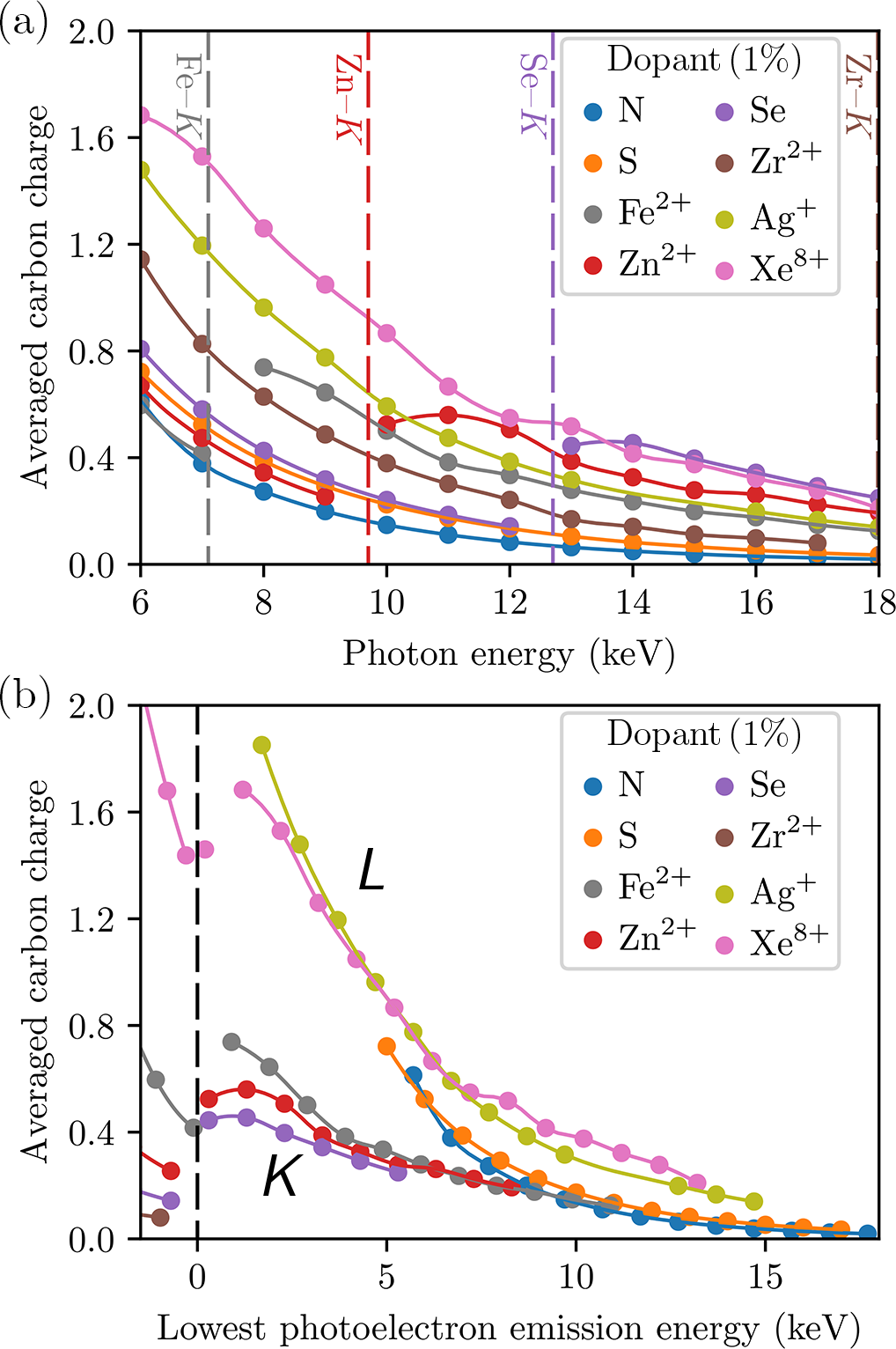}
\caption{Relationship between light-atom ionization and the energy of photoelectrons ejected by trace heavy elements. Each trace corresponds to a target of composition C$_{613}$N$_{193}$O$_{185}$X$_{10}$, where X is the dopant denoted in the legend. (a) shows the intensity-averaged charge of carbon against the photon energy of the pulse. The points are shifted in (b) to align with the separation between the deepest ionizable shell (DIS) of the dopants at 18 keV and the photon energy, representing the lowest-energy photoelectrons emitted by the DIS (when positive). While the severity of ionization varies considerably between targets at any given photon energy, the traces in (b) form two distinct groups according to the shell number of the DIS (annotated). The interpolating lines are included as a guide for the eye. Photon fluence was fixed at $10^{12}$ ph$\cdot$\si{\micro\metre}$^{-2}$, using a 15 fs FWHM Gaussian profile. Simulations were run from -18 fs to +18 fs relative to the pulse peak.}  
\label{fig:anti-goldilocks}
\end{figure}

\begin{figure*}
\includegraphics[width=15cm]{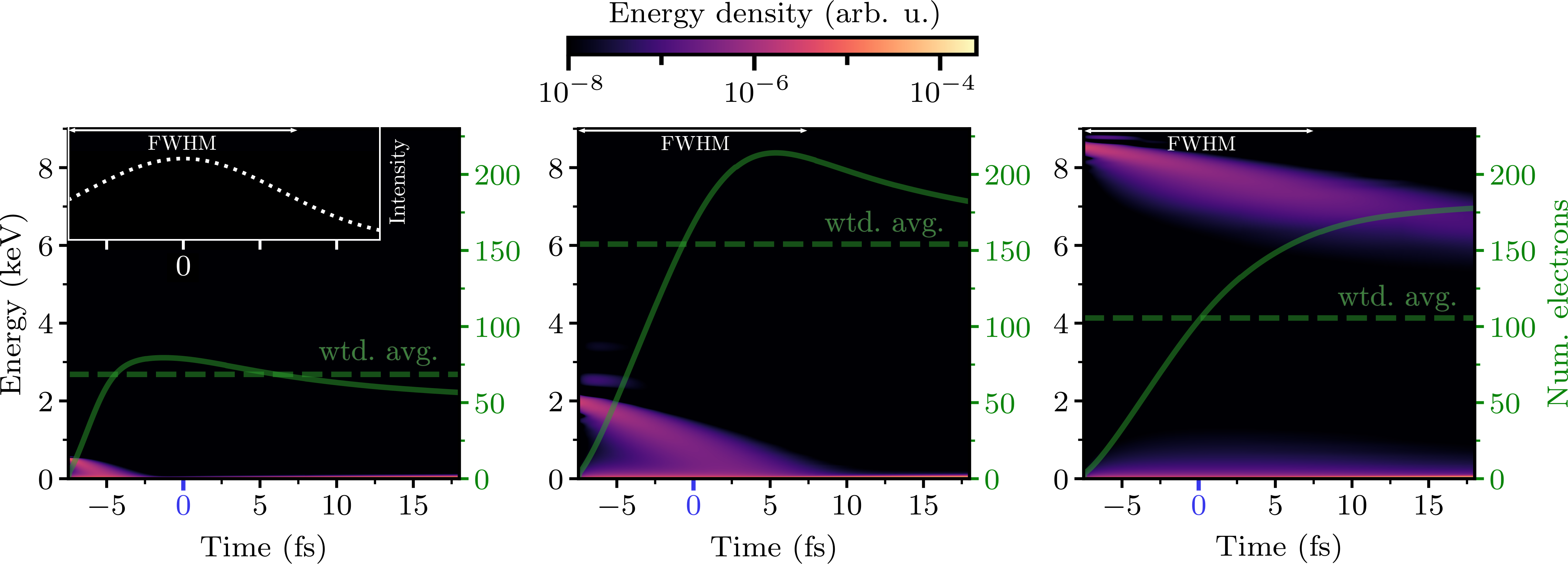}
\caption{Evolution of electron-ionization cascades under plasma conditions in a light-atom target. Each plot shows a cascade initiated by a free electron of energy (a) 0.5 keV (b) 2 keV (c) 8.5 keV. Each cascade occurs during a full dynamical simulation of the C$_{613}$N$_{203}$O$_{185}$ target under the 15 fs FWHM 9 keV pulse, in equivalent simulation conditions to Fig.~\ref{fig:anti-goldilocks}, with the cascade initiated when the pulse first reaches the FWHM intensity (-7.5 fs). The right axis of each plot shows the density of the cascade, normalized to correspond to the number of electrons. Dashed horizontal lines correspond to the intensity-averaged electron count. The 500 eV cascade (oxygen Auger electron) deposits a small amount of energy quickly, while the 8.5 keV cascade (oxygen photoelectron) is largely outrun by the pulse. The 2 keV cascade is the most damaging, depositing its energy over a timescale similar to the 15 fs FWHM of the XFEL pulse. Each cascade was tracked by injecting a negligibly small electron density into the system at $t=-7.5$ fs. At late times, the cascade densities are reduced by recombination. } 
\label{fig:single_cascade}
\end{figure*}

\subsection{Photon and primary electron energy} \label{subsec:cascade}

Fig.~\ref{fig:anti-goldilocks}(a) shows that the choice of dopant substantially impacts global ionization. At all energies, the intensity-averaged carbon charge varies widely across the targets. Most traces show decreasing damage with increasing photon energy, consistent with prior work~\cite{neutze}; however, the traces for the targets doped with Fe$^{2+}$ ($Z$=26), Zn$^{2+}$ ($Z$=30), and Se ($Z$=34) buck this trend, with the transition over the K-edge of the dopant increasing the average charge by roughly a factor of 2 or more in each case. This reflects the substantial contribution of the K-shell to the primary ionization of these dopants; the 1s orbitals dominate the photoabsorption cross-sections of their atoms and are replenished on a subfemtosecond timescale by Auger cycling (see Sec.~\ref{subsec:ionization_modes}).

Perhaps more surprisingly, the intensity-averaged carbon charge \textit{continues} to increase up to $\sim$\;2~$\si{\kilo\eV}$ above an inner-shell absorption edge, despite a decreasing photoabsorption cross-section. This is most clearly seen in Fig.~\ref{fig:anti-goldilocks}(b). %
Since the separation from the edge corresponds to the energy of the photoelectrons emitted from the shell, such a result might be explained if the photoelectrons of $\sim$2 keV are unusually damaging. Thus to isolate the source of the effect, it is necessary to quantify the damage that a single primary electron emission induces.
We add an artificial, low-density primary-electron peak to the continuum in the simulation of the CNO (C$_{613}$N$_{203}$O$_{185}$) when the pulse first reaches the FWHM intensity, then track the resulting electron cascade as in Sec.~\ref{subsec:ionization_modes} (see Eq.~\eqref{eq:separate_pop})

Fig.~\ref{fig:single_cascade} confirms that the intensity-averaged ionization caused by cascades initiated by 2 keV 
primary electrons is higher than those caused by 0.5 keV \textit{and} by 8.5 keV primary electrons. This observation is consistent with the local maximum of ionization near the photoelectron emission energy of 2 keV in Fig.~\ref{fig:anti-goldilocks}(b).

\begin{figure}
    \includegraphics[width=2.8in]{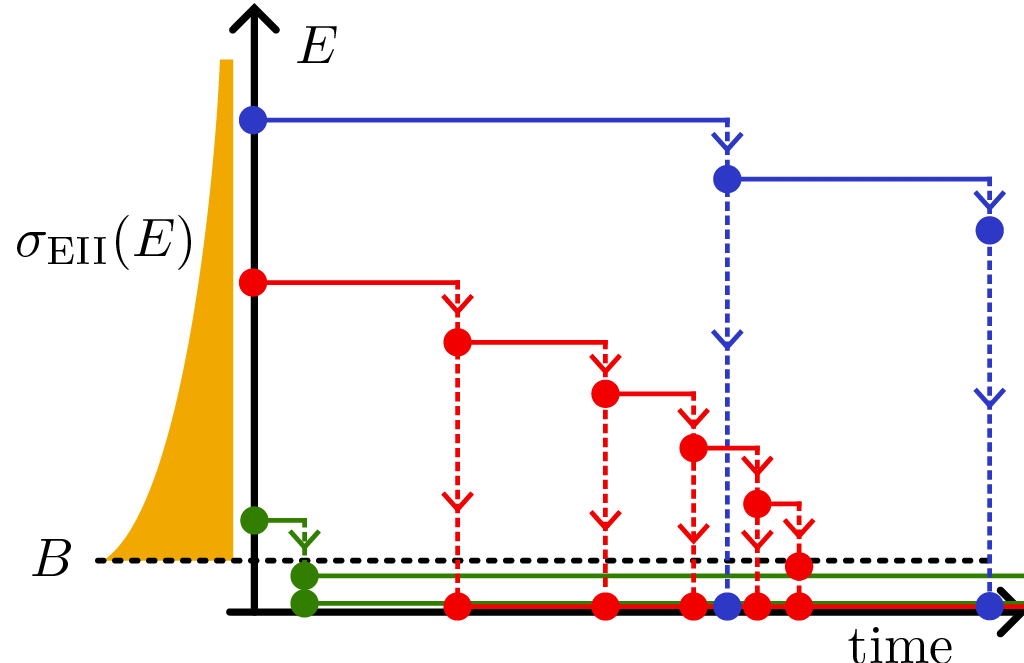}
    \caption{Diagram illustrating how intermediate-energy free electrons can be more damaging on sufficiently short timescales. Three free electrons, represented as dots, are shown initially with low (green) intermediate (red) and high (blue) energies incident on a gas of bound electrons with binding energy $B$. The horizontal lines represent the mean time between electron-impact ionization (EII) events, which is smaller for lower energy free electrons. Each branching event represents the most likely EII process: an electron of energy $E$ `only just' ionizes an atom, leaving one electron at zero energy and another at $E-B$. Because the EII cross-section ($\sigma_\text{EII}$, indicated by the yellow trace) is lower for faster free electrons, there exists a window of time where, on average, the intermediate-energy electron cascade will have ionized more atoms than the high- and low-energy cascades.}
    \label{fig:cascade_sketch}
\end{figure}

To understand why the intermediate-energy cascade is more damaging, one can consider a free electron of energy $E$ interacting with a gas of hydrogenic atoms, with electrons bound by energy $B$. As $E$ falls, the EII cross-section grows, causing an increased ionization rate down to $E \simeq B$~\cite{kimBinaryencounterdipoleModelElectronimpact1994,suno}. This creates a trade-off: higher-energy cascades progress more slowly, but have the potential to free more electrons. Indeed, this effect can be seen in comparisons of the evolution of electron cascades at various energies in neutral targets presented in prior work~\cite{chapman2014,caleman2015}. 
When counting all ionization events over a fixed duration, there will be a maximally ionizing cascade energy. (Fig.~\ref{fig:cascade_sketch} gives a qualitative sketch of this idea.) This implies the existence of a `maximally damaging' cascade energy $E_{max}$, loosely corresponding to the energy that maximizes the intensity-averaged ionization. Under conditions that induce significant atomic disorder during the pulse, $E_{max}$ is likely lowered by Bragg gating~\cite{barty2012,caleman2015,chapman2014}. 
Due to the higher photoabsorption cross-section near the edge, the most damaging \textit{photons} eject primary electrons with energies below $E_{max}$; however, simulations that control for the rate of primary ionization show the difference to be slight (about 0.5 keV, see Appendix~\ref{app:isolated_energy}). %

Fig.~\ref{fig:anti-goldilocks}(b) plots the average carbon charge against the lowest photoelectron energy (LPE) at emission instead of the photon energy. The traces form two distinct groups: the `K-group', consisting of targets doped by an element whose `deepest ionizable shell' (DIS) in the considered energy range is the K-shell (N, S, Fe, Zn, Se), and the `L-group', where it is the L-shell (Ag, Xe). The split in groups again appears to be a result of Auger cycling. As the DIS of each heavy dopant is maintained near maximum occupancy by decay processes, the production rate of ionizing electrons by Ag$^+$ and Xe$^{8+}$ is scaled by a factor of $\sim$\;4 relative to the other dopants at an equivalent LPE, due to the higher electron capacity of their L-shells.  

In general, the variation in electronic damage across the targets with respect to photon frequency is predominantly controlled by the LPE of the dopants and the shell number of the DIS. However, the traces of the N, S, and Fe-doped targets deviate from the rest of the K-group in Fig.~\ref{fig:anti-goldilocks}(b) for photon energies at or below 8 keV. In these cases, the photoelectrons ejected from the light atoms are of a low enough energy to contribute significantly to their secondary ionization (relative to the dopants). Since the Zn and Se K-edges are above 8 keV, their traces show no deviation. It is unsurprising that the L-group does not contain such outliers, as the primary electron contribution of the L-group dopants is much more significant.

\subsection{Damage landscape} 
\label{sec:damage_landscape}

\begin{figure}[H]
\centering
\makebox[\linewidth][c]{
\hspace{40pt}\includegraphics[width=7.24409436834in]{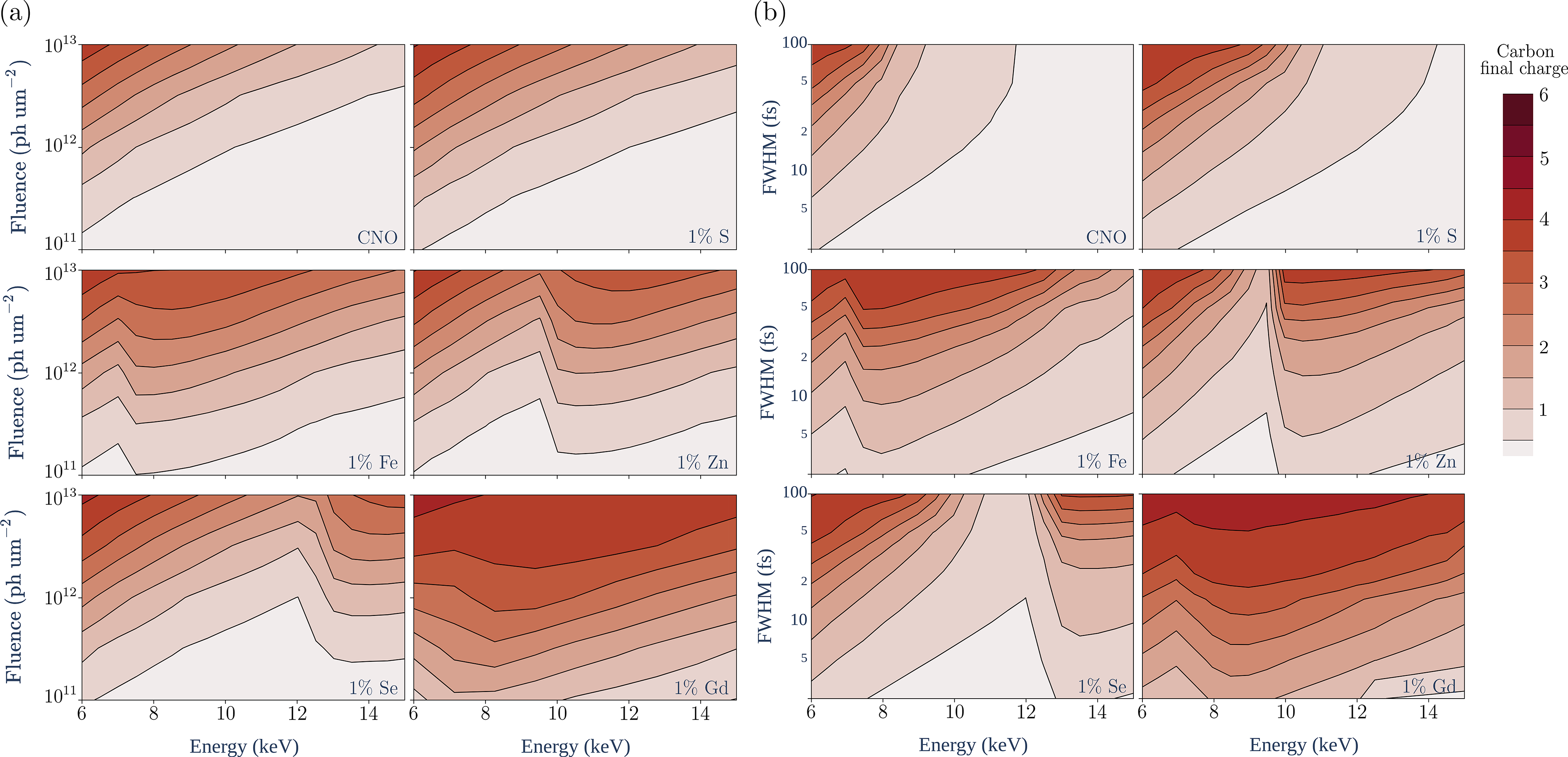}
}
\caption{Effect of chemical composition on the electronic damage landscape. (a) 15 fs FWHM Gaussian pulses. (b) Gaussian pulses of $10^{12}$ ph$\cdot$\si{\micro\metre}$^{-2}$ fluence. Each plot maps the average charge of carbons at the end of the illumination (+1.2 FWHM) to a target of composition C$_{613}$N$_{193}$O$_{185}$X$_{10}$, where X is the element denoted in the lower-right corner. Each plot uses (a) 133 or (b) 190 data points. The sharp features in the plots for targets doped by heavier elements correspond to the dopant absorption edges.}
\label{fig:damage_landscape}
\end{figure}

Repeating the simulations shown in Fig.~\ref{fig:anti-goldilocks} while ignoring heavy-atom secondary ionization had little effect on the traces, as predicted by the analysis in Sec.~\ref{subsec:ionization_modes}. Since this substantially reduces the number of configurations to be processed in the EII and TBR calculations, additional simulations were performed using this approximation to map the landscape of damage \cite{neutze} for each target (Fig.~\ref{fig:damage_landscape}).

Contour plots of the charge accumulated by the carbon atoms, as functions of photon energy and fluence (Fig.~\ref{fig:damage_landscape}(a)) or pulse width (Fig.~\ref{fig:damage_landscape}(b)), show significant differences due to the dopants.
The effect of sulfur on the dynamics is relatively consistent; across the parameter space of Fig.~\ref{fig:damage_landscape}, the S-doped target experiences levels of ionization that the pure CNO target only sees at 20--40\% higher intensities.  
The impact of the heavier dopants is more variable due to their deep absorption edges, which are distinctly visible in these plots -- the most extreme case is the Se-doped target, where the elevation in damage from increasing the photon energy from 12.5 keV to 13.5 keV is similar to the elevation in damage from increasing the pulse fluence by a factor of 3--4 (Fig.~\ref{fig:damage_landscape}(a)), or from increasing the pulse width by an order of magnitude (Fig.~\ref{fig:damage_landscape}(b)).  

Inspection of Fig.~\ref{fig:damage_landscape} shows that the ionization in the doped targets consistently sees a local maximum around a LPE of 2 keV, suggesting this may serve as an approximate value for $E_{max}$ in this regime (see Sec.~\ref{subsec:cascade}). Both the slow Auger electrons and fast photoelectrons ejected by light atoms are well away from this value for $E_{max}$ in typical hard X-ray SFX experiments, while the primary electrons ejected by heavy atoms are almost always closer. Indeed, this is the case in the lysozyme.Gd scenario considered in Sec.~\ref{sec:lysozymeGd}. For example, the $\sim 4$ keV photoelectron peak of Cl, observable in Fig.~\ref{fig:free_bound_combined}(a), is closer to 2 keV than the $> 6.5$ keV light-atom photoelectrons and thus more damaging, and the same is true of the Auger electrons, which are typically $\sim2.3$ keV for Cl and $<500$ eV for light atoms. The tendency of heavy atoms to eject electrons with `intermediate' energies closer to $E_{max}$ thus contributes to the strength of their effect.

Fig.~\ref{fig:damage_landscape} allows for the magnitude of damage in targets containing different heavy elements to be compared on the keV energy scale. This picture is only accurate in a coarse-grained sense, as resonance-related phenomena near absorption edges that modulate the fine structure of the damage landscape~\cite{kanterHiddenResonances2011, RudekResonantHeavyAtomIonization2012, rudekRelativisticResonantEffects2018, hoRoleTransientResonances2020} are not modeled here.
In preparation of these diagrams, we were careful to avoid any photon energies close to resonance. 
We neglected ionization potential depression (IPD)~\cite{vinkoDensityFunctionalTheory2014,vinkoInvestigationFemtosecondCollisional2015,ciricostaMeasurementsContinuumLowering2016,royleKineticModelingXray2017}, which a short order of magnitude calculation shows to be limited to $\sim$50eV at biomolecular densities.

\onecolumngrid

\twocolumngrid

\section{Discussion} \label{sec:discussion}

The results presented here suggest the absorbed dose and evolution of the electronic structure are strongly affected by the presence of heavy elements. For all targets considered, a high-energy regime exists in which the general principle that `higher photon energy = less damage' remains true. However, for certain targets, the maximum photon energies available from XFEL sources are too low to enter this regime (Fig.~\ref{fig:damage_landscape}). This is a departure from conventional crystallography, where higher energies are broadly beneficial. More positively, this suggests SFX experiments may be able to control the severity of damage through the choice of X-ray frequency or solvent composition. For example, excluding salt from the solvent in the simulation of non-derivative lysozyme  (Fig.~\ref{fig:free_bound_combined}) reduced the average carbon charge at $t=10$ fs from +1.53 to +1.28.

Elements much heavier than sulfur, such as Se and Gd, are often introduced to targets for anomalous phasing, generally in tandem with an X-ray energy just above their ionization edges \cite{hunter2016,gorel2017TwoColour,galli,sonDeterminationMultiwavelengthAnomalous2013}. It is commonly held that difficulties in applying the technique to SFX can be mitigated by using increased heavy atom concentrations \cite{hunter2016,gorel2017TwoColour}, but the results of the present work suggest this is a trade-off in the ultrafast regime, boosting the anomalous signal at the cost of additional damage to the light atom structure. For example, consider the effect of phasing with the Se K-edge on global damage. For the Se-doped target, Fig.~\ref{fig:damage_landscape}(b) shows adjusting the X-ray energy from 14 keV to 12 keV is as effective at reducing the ionization of light atoms as compressing the pulse width from 50 fs to 5 fs. Alternatively, this cost might be reduced if phasing is performed with an X-ray frequency well above the absorption edge, as has been done in native phasing experiments \cite{nakane2015,galliSulfurExperiment}.

There is preliminary evidence that these considerations are relevant in practice. Nass \textit{et al.} attributed unusual noise in the scattering profile of lysozyme.Gd nanocrystals, absent from thaumatin nanocrystals to increased radiation damage induced by Gd~\cite{nassStructuralDynamicsProteins2020}.
Our modeling supports this inference, as it found the addition of just Gd to the target (alongside C, N, O, and S) increases the pump-pulse intensity-averaged charge of C, N, O by 25.2\%, 26.9\%, 26.7\%, respectively.
However, Fig.~\ref{fig:damage_landscape} suggests that, at the fluence of this experiment (nominally $3.5 \times 10^{12}$ ph$\cdot$\si{\micro\metre}$^{-2}$, though likely lower in reality~\cite{nassStructuralDynamicsProteins2020,naglerFluenceError2017}), the 7.1 keV photon energy used (just low enough to avoid ionizing the L$_3$-edge of the Gd ions) would be close to the optimal choice for minimizing damage in this case. 
Indeed, for the $1.75 \times 10^{12}$ ph$\cdot$\si{\micro\metre}$^{-2}$ fluence pulse simulated in Sec.~\ref{sec:lysozymeGd}, increasing the photon energy to 9 keV caused the intensity-averaged charge of protein light atoms to rise from +0.72 to +0.74 for lysozyme.Gd in 0.1 M Gd. For reference, it fell from +0.57 to +0.34 for lysozyme in water.

The obvious limitation of the presented model is its zero-dimensional treatment. Target substructures such as metal cofactors often have order 10 nm separations, and the large-scale distribution of heavy atoms is generally non-uniform due to the differing compositions of the protein and its aqueous environment. Naively, this suggests heavy atoms produce a `sphere' of electronic damage in their local region, with the distance spanned dependent on photon energy.  However, whether such heterogeneous correlations actually occur is complicated by the non-uniformity in the large-scale solvent-protein structure of real targets. A model fit for exploring this possibility will likely need to break the crystal symmetry and account for spatial variation in electron density on the global scale of the target. Additionally, the effect of heavy atoms suggests a significance to the mother liquor composition in conjunction with electron transfer across the crystal boundary. It is likely, for example, that the high-energy electrons originating in the mother liquor replace those of the crystal to an extent dependent on the crystal size~\cite{caleman2011}.

Finally, we remark on the impact of the pulse profile on these dynamics. Previous works have generally modeled the temporal intensity profile as either square~\cite{hau-riegeNonequilibriumElectronDynamics2013,royleKineticModelingXray2017} or Gaussian~\cite{royleKineticModelingXray2017,Ren2023NonLTE-CCFLY}. The ionization dynamics proved to be sensitive to this choice (see Fig.~\ref{fig:pulse_profile}). Repeating the lysozyme.Gd simulation (Fig.~\ref{fig:free_bound_combined}) with a square pulse of the same fluence, energy, and FWHM resulted in light-element intensity-averaged charges that were 19\% lower than under the Gaussian pulse. This can be attributed to an outsized effect by the earlier EII cascades~\cite{jonssonPulseProfile2015} induced by the Gaussian pulse during its leading tail. Such cascades have a long period to ionize the target before the bulk of the elastic X-ray scattering. This indicates a necessity for modeling of the dynamics under more realistic SASE pulse profile statistics.

\section{Conclusions}

The zero-dimensional non-Maxwellian model employed in this study suggests that a significant amount of damage to biological macromolecules under XFEL illumination is seeded by heavy atoms, with even the presence of native sulfur atoms significantly elevating global ionization. This result might appear surprising given such targets only contain heavy elements in trace quantities; however, closer inspection shows this outcome to reasonably follow from two key points: (I) Heavier species emit photo- and Auger electrons at much higher rates, considerably boosting the number of secondary ionization cascades instigated within the light atom bulk. (II) Relative to the 10 fs timescale on which the structural signal is captured in XFEL experiments, light-atom-sourced electron avalanches will either have a very low energy and thus dissipate prematurely, or have a very high energy and thus a small EII rate; in contrast, avalanches initiated by heavy atoms, with energies between these two extremes, more severely degrade the captured structural signal. The non-Maxwellian treatment of the electron continuum was necessary to capture this latter point.

The addition of heavy atoms to the environment of proteins -- such as potassium and sodium ions in the mother liquor -- is routine in protein crystallography; however, the results of this work suggest that on the short timescales of XFEL pulses their use becomes a trade-off for additional ionization. Judicious choices to reduce the number of low--intermediate energy primary electron emissions may thus improve experimental outcomes where damage is a concern, or where controlling for damage across pulse parameters is necessary. Specific to \textit{de novo} refinement, anomalous phasing methodologies that allow for weaker anomalous signals would see a reduction in damage-induced noise, suggesting a strength for native phasing over artificial introduction of heavier elements such as Gd. Further, the production of primary electrons near the maximally ionizing energy can be avoided entirely with careful choice of photon frequency.

This work has restricted its broader analysis of the electronic damage landscape to targets where heavy elements make up 1\% of the atomic population, but this is far from sufficient to generalize the influence of heavy atoms across the varied ratios seen in real targets, including targets containing multiple heavy elements. However, it is evident that experimental differences generally considered marginal in traditional crystallography can substantially affect the amount of radiation damage suffered by targets in the ultrafast regime. This complexity emphasizes the value of using theoretical modeling to inform SFX experimental design -- a role it is already fulfilling \cite{dickersonRADDOSE-XFEL2020} -- particularly as a tool for gauging the viability of successful refinement in the high-intensity regime. The zero-dimensional framework employed in \texttt{AC4DC} can capably examine the complete electronic damage dynamics across a large number of candidate pulse parameterizations without significant investment of computational resources. For studies concerned with the ions' motions, the simulation may be integrated within a hybrid plasma-MD framework~\cite{kozlov2020_model,dawodMolDStruct2024}, where delegation of the ultrafast electron dynamics to a zero-dimensional model makes simulating the molecular dynamics of 10--100 nm scale structures feasible.

\newcommand{\sectionstar}[1]{%
  {\setcounter{secnumdepth}{0}\section{#1}\setcounter{secnumdepth}{3}}%
}

\sectionstar{Acknowledgements}

We thank A/Prof. Nadia Zatsepin, Swinburne University of Technology, Hawthorn, Australia, for feedback on the discussions of SFX experimental methods and Thippayawis Cheunchitra, University of Melbourne, Parkville, Australia, for inspiring examination of the damage landscape above deep absorption edges.

\sectionstar{Data availability}

Simulation input files for Sec.~\ref{sec:lysozymeGd} may be found at \url{https://github.com/Phoelionix/AC4DC/tree/master/input/lysozyme.} Input files for Sec.~\ref{sec:deep_absorption_edges} may be found at \url{https://github.com/Phoelionix/AC4DC/tree/master/input/templates/doped_targets}. Additional data used in this work is available upon reasonable request to the corresponding author.

\appendix

\section{Numerical method}
\label{app:numerics}

\subsection{Cubic spline representation of the free-electron energy distribution}

The collision kernel on the right-hand side of the Boltzmann equation, Equation (\ref{eq:Boltzmann}), takes the form of a sum over distinct processes,

\begin{align}
    \mathcal{Q}[P_\xi, f](\epsilon) =& 
    \mathcal{Q}^{\rm Photo}[P_\xi](\epsilon) 
    + \mathcal{Q}^{\rm Auger}[P_\xi](\epsilon)
    +\mathcal{Q}^{\rm EII}[P_\xi,f](\epsilon)\nonumber \\
    &+\mathcal{Q}^{\rm TBR}[P_\xi,f](\epsilon)
    +\mathcal{Q}^{\rm EE}[f](\epsilon)
    \label{eq:Qee_explicit_appendix}
\end{align}

The solution of these coupled rate equations presents several numerical challenges:
\begin{enumerate}
    \item The primary-ionization terms $\mathcal{Q}^{\rm Photo}$ and $\mathcal{Q}^{\rm Auger}$ are essentially Dirac delta-like source terms; such singularities often destabilize numerical PDE solutions.
    \item The number of possible electron configurations scales factorially with the number of electrons in a given atom.
    \item The three-body recombination term $\mathcal{Q}^{TBR}$ is quadratic in the free-electron energy distribution $f$, leading to worst-case $N^3$ complexity if the representation of $f$ is $N$ dimensional.
    \item The electron-electron recombination term $Q^{EE}$ depends on derivatives of $f$.
\end{enumerate}

We seek a representation for $f$ that i) is inherently smooth and at least once differentiable, ii) is capable of representing strongly-peaked ionization functions without Gibbs-like phenomena, and iii) admits a computationally efficient representation of $\mathcal{Q}^{TBR}$. 

The standard approach~\cite{morganELENDIFTimedependentBoltzmann1990} to non-Maxwellian plasma simulation solves the Boltzmann
equation using finite differencing of $f(\epsilon, t)$. We take a more general
approach, expanding $f$ with respect to
a time-invariant basis $\mathcal{B} = \{ \phi_i(\epsilon), i=1...N \} $,
contracted with time-varying expansion coefficients $c^j(t)$ and multiplied by
an explicit weight function $w(\epsilon)$,
\begin{equation}
	f(\epsilon, t) = w(\epsilon)\sum_i c^i(t) \phi_i (\epsilon) \; .
\end{equation}
This expansion must be valid both at early times, when $f$ resembles a sum of Dirac delta functions at the photoelectron and Auger electron energies, and at later times, when it resembles a Maxwell-Boltzmann distribution $f_T(\epsilon)
\sim \sqrt{\epsilon} \exp(-\epsilon/k_B T)$. Typical orthogonal bases for $L^2(\mathbb{R})$ (e.g. Legendre polynomials, Fourier sines) are simple to differentiate and suitable for describing the smooth features of a Maxwellian, but suffer from Gibbs' phenomena in the vicinity of the strongly peaked atomic lines. The other typical basis choice -- the orthogonal rectangles of finite-element analysis -- works well when describing sharp peaks, but is susceptible to numerical instabilities when calculating derivatives \cite{khoRelaxationSystemCharged1985}.

Order-$k$ B-splines ($k=3$ in this work) are a good compromise between the two approaches. 
A collection of $N$ splines of order $k$ is defined over a grid of non-decreasing `knot points' that collectively form the \textit{knot vector} $\bm{T} = \{ t_j | j=0...N+k-1,
t_k \ge t_k-1\}$ \cite{boorPracticalGuideSplines1978}.
An expansion $f$ constructed from order-$k$ B-splines is automatically
polynomial away from the knots, and at a knot point of multiplicity $m$ has
$k-1-m$ continuous derivatives. The knot vector is constructed such that
the first and last knots have multiplicity $k$, while all interior points are
distinct. With this choice of knot vector, the $\phi_k$ form a partition of unity -- $\forall x \in
[t_0, t_N+k-1], \sum_n \phi_n(x) = 1$.

Such functions are not mutually orthogonal, possessing a non-trivial overlap matrix $S_{ij} := \int d\epsilon w(\epsilon) \phi_i(\epsilon)
\phi_j(\epsilon)$. This basis permits an approximate representation of
a narrow primary electron peak by a $C^1$ piecewise polynomial, ensuring that
the distribution has a well-defined derivative everywhere.
Eq.~\eqref{eq:Boltzmann} may then be recast in a finite form by integrating both sides against a test basis element $\phi_j$,

\begin{align}
	\int d\epsilon \phi_j(\epsilon) \frac{\partial f(\epsilon, t)}{\partial t} &=
	\int d\epsilon \phi_j(\epsilon) \mathcal{Q}[\bm{P}, f](\epsilon, t) \\
	\Rightarrow S_{ij} \frac{d c^i(t)}{\partial t} &= Q_j [\bm{P}, f](\epsilon, t)\ ,
\end{align}

which has the added benefit of rendering the $Q$ tensors sparse.

In our approximation scheme, we assume
\begin{enumerate}
    \item Purely (semi)classical collisional dynamics,
    \item Independent, decoupled atoms,
    \item Spatial homogeneity.
\end{enumerate}

Assumption 2) is equivalent to discarding second-order and higher correlations in the electrons' distribution function.

We separated the time-dependent energy distribution of free electrons $f(\epsilon, t)$ from the discrete probability distribution $P_\xi (t)$ capturing the classical populations of each atomic state $\xi$.
The Boltzmann and master equations \eqref{eq:Boltzmann} and \eqref{eq:Master}
then generate a deterministic time evolution for the classical energy-state distribution of the electrons.

Further numerical methods employed included: (I) An asynchronous implicit-explicit (IMEX) method, stepping the stiff but computationally cheap free-electron interactions with much shorter steps than the bound-bound and bound-free contributions. (II) Adaptive time steps to avoid divergence in $f(\epsilon,t)$.

\subsection{Explicit form of numerical couplings}

We adopt the following notation:
\begin{description}
    \item[$a$] Index for atom types, e.g. C,N,O,...
    \item[$\xi,\eta$] Indices for electron configurations of a particular atom type
    \item[$\Gamma_{\xi\to\eta}$] Transition rate from configuration $\xi$ to configuration $\eta$. Square brackets indicate an $f$-dependent decay rate.
    \item[$B_{\xi\eta}$] Difference in total binding energy between electron configurations, $B_\xi-B_\eta$
\end{description}

\subsubsection{Electron-Electron interactions}

We model the Coulomb-mediated relaxation of the electron gas using a standard Fokker-Planck kernel, specifically the form quoted by Ref.~\cite{khoRelaxationSystemCharged1985},

\begin{equation}
	\mathcal{Q}^{ee}[f](\epsilon) = -\frac{\partial J^{ee} [f](\epsilon)}{\partial \epsilon},
	\label{eq:Qee}
\end{equation}
where the electron current $J^{ee}$ has the form
\begin{align}
	J^{ee}(\epsilon) &= \alpha \left[ F(\epsilon, t) \left(\frac{f}{2\epsilon}
	- \frac{\partial f}{\partial \epsilon} \right) - G(\epsilon, t) f \right] ,
	\label{eq:Jee}\\
			F(\epsilon, t) &= 2\epsilon^{-1/2} \int_0^\epsilon x f(x, t)dx
			+ 2 \epsilon \int_\epsilon^\infty x^{-1/2} f(x, t) dx,\\
			G(\epsilon, t) &= 3\epsilon^{-1/2} \int_0^\epsilon f(x, t) dx,\\
			\alpha&=\frac{2}{3} \pi e^{4}(2 / m)^{1 / 2} \ln \Lambda . 
\end{align}

Notice that the electron-electron interaction strength requires an estimate of the Coulomb logarithm $\ln \Lambda = \int d\chi/\chi$~\cite{morganELENDIFTimedependentBoltzmann1990,royleKineticModelingXray2017,oxeniusKineticTheoryParticles1986}. In our case, we cut the integral off using the Debye length and radius of closest approach 
$r_{0}
= e^2/4\pi\varepsilon_0 k_B T$, giving $\Lambda = n 4\pi{\lambda_D}^3$ ~\cite{oxeniusKineticTheoryParticles1986}. 
The temperature $T$ here represents the notional temperature of the thermalized free electrons, estimated by a dynamical fit to the Maxwellian portion of the distribution.

\subsubsection{Photoionization and Auger decay}

Both processes produce primary electron emission spectra with very narrow linewidths. For numerical stability, we artificially broaden these features  into a linear combination of the two basis splines nearest the desired energy, with coefficients chosen such that
\begin{align} 
	\int d\epsilon \Delta_E(\epsilon) = 1, & & \int d\epsilon \Delta_E(\epsilon) \epsilon = E .
\end{align}
This approach avoids introducing spurious negative-density parts to the
distribution, as would be the case if a naive expansion of $\delta(E-\epsilon)$ were used. This should be a mild approximation in the XFEL regime, as the semi-empirical estimate of Ref.~\cite{khoRelaxationSystemCharged1985} indicates such broadening occurs on a subfemtosecond timescale.
 
\begin{align}
	\mathcal{Q}_a^{\text{Pht}}[{P}_a](\epsilon, t) &= 
 \sum_{\xi \in \mathcal{C}_a} P_\xi(t)
	\sum_{\eta \neq \xi} \Delta_{\hbar\omega-B_{\xi
	\eta}}(\epsilon)\Gamma^\text{Pht}_{\xi \to \eta}\\
 	\mathcal{Q}_a^{\text{Auger}}[{P}_a](\epsilon, t) &= \sum_{\xi \in \mathcal{C}_a} P_\xi(t)
	\sum_{\eta \neq \xi} \Delta_{B_{\xi\eta}}(\epsilon)\Gamma^\text{Auger}_{\xi \to \eta}
\end{align}

\subsubsection{bound-free interactions}

Expressions for $\mathcal{Q}_a^{\text{EII}}$ and $\mathcal{Q}_a^{\text{TBR}}$ were derived according to the work of Ref.~\cite{leonovTimeDependenceXray2014}, to which we direct the reader.

\subsection{Dynamic grid implementation}  

The cubic spline treatment of the plasma code has previously been shown to well-approximate the evolution of relatively static Boltzmann-governed systems with as few as 10 energy grid points/knots \cite{khurana}. However, this requirement grew by an order of magnitude when the initial distribution of particles was modestly (relative to the energy scale of the present work) displaced from equilibrium with a thermal bath. The unbound electron continuum of a biological target under XFEL illumination sees an acute difference between its state at early times and its partially equilibrated state near the end of the pulse. Moreover, these systems see two complications not present in the systems considered by Khurana \textit{et al.}: (I) The narrow, low-energy peak of the Maxwellian at early times is difficult to fit due to the rigid energy conservation condition of $\frac{d^2f}{d\epsilon^2}\approx0$. (II) The thermal bath is substituted for sharp, high-energy emission profiles, which, like the Maxwellian, shift significantly over the course of the pulse as they relax through electron interactions~\cite{hau-riegeNonequilibriumElectronDynamics2013}. 

 An adaptive grid was implemented to address these issues. A set of static low-density regions spans the full energy range and a set of \textit{dynamic} high-density regions spans i) the Maxwellian and ii) up to four of the most dominant high-energy peaks. We define a set of energy ranges (regions), each with an associated knot density function $\xi(\epsilon)$ that is only non-zero within the region. The local knot density of the full grid $\Xi(\epsilon)$ is then defined as
\begin{equation}
\Xi(\epsilon) = \max(\xi_1,\xi_2,...,\xi_{n-1},\xi_{n}) .
\end{equation}

The high-density regions of the electron continuum are dynamically updated throughout the simulation to follow their respective features. A high density of knots supports the sharp features in the continuum at early times, then redistribute to continue to support the growing and shifting peaks as the electron population equilibrates. Testing showed accurate fitting of the continuum at early times to be particularly crucial, so a partial run of the simulation with a `guess grid' is used to obtain the initial grid for the actual simulation. This flexible approach drastically reduced the number of knots necessary to achieve convergence in the evolution of the ionic states. 

The spline basis is transformed with a 64-point Gaussian quadrature. Transformations in the simulations of this work were all performed an order of 10 times. Testing found the convergence to be independent of the associated error for $>$100 transformations. The photoelectron peaks were identified dynamically based on their maximum energy density relative to the transition energy region, without regard for the prior basis.

\section{Influence of cascade energy with controlled seeding rate} \label{app:isolated_energy}

\begin{figure}[H]
    \centering
    \includegraphics[]{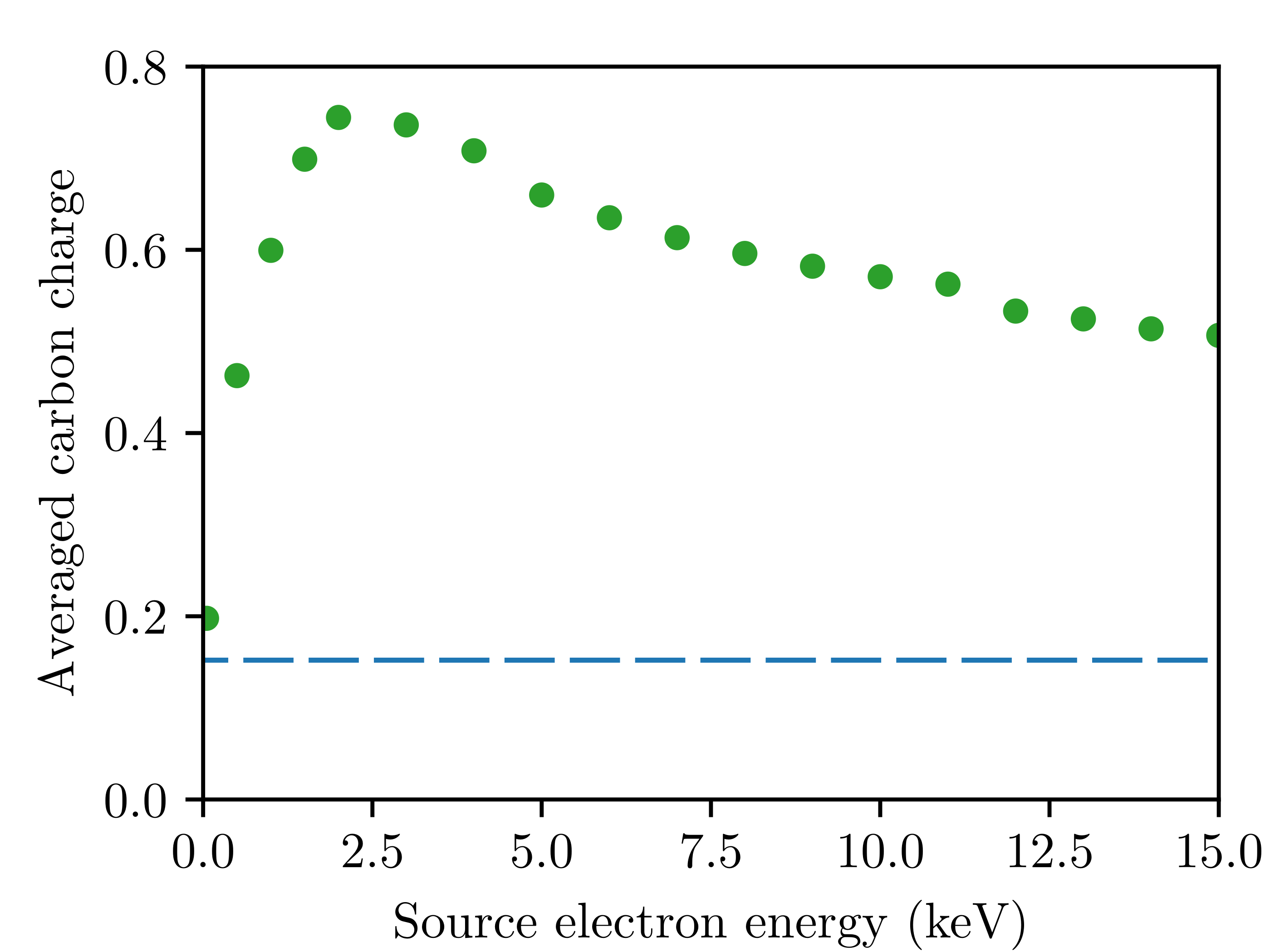}
    \caption{
        Relationship between the energy of electrons seeded within a solid light-atom target, C$_{613}$N$_{203}$O$_{185}$, and the intensity-averaged charge of carbon. Each dot corresponds to a simulation of the light-atom control subjected to a $10^{12}$ 10 keV ph$\cdot$\si{\micro\metre}$^{-2}$, 15.0 fs FWHM pulse, but with an artificial injection of free electrons at the energy denoted by the horizontal axis. Electrons were injected at the same rate across all simulations, proportional to the intensity. Specifically, electrons were added at three times the photoionization rate of the undamaged target at the given intensity. The dashed line indicates the averaged carbon charge when no electrons are injected.}%
    \label{fig:goldilocks_electron_injection}
\end{figure}

Increasing the X-ray energy further above an absorption edge raises the energy of the seeded EII cascades. At the same time, the number of these cascades decreases due to the reduced absorption cross-section at higher energies. To isolate the relationship between the energy of a primary electron and the damage caused by the resulting cascade, \texttt{AC4DC} was modified to model the ionization of the light-atom control with additional electrons injected at a rate independent of their energy. Each system was simulated under a 15.0 fs FWHM Gaussian pulse with a fluence of $10^{12}$ 10 keV ph$\cdot$\si{\micro\metre}$^{-2}$. The injection rate was fixed at three times the photoionization rate of the \textit{undamaged} target for a given incident intensity, coarsely corresponding to the ionization rate of the Zn-doped target by its Zn ions under the same pulse conditions. (Zn was chosen due to the distinct local maximum it produces, as shown in Fig.~\ref{fig:anti-goldilocks}.) The resulting relationship between the electron source energy and carbon ionization (Fig.~\ref{fig:goldilocks_electron_injection}) shows strong similarities to the relationship between LPE and carbon ionization in the Zn-doped target shown in Fig.~\ref{fig:anti-goldilocks}(b), including a similar position and magnitude for the local maximum.

\section{Model comparisons} \label{app:comparisons}

\subsection{Amorphous carbon -- \texttt{ddcMD}}

\begin{figure}[H]
\includegraphics[width = 8.5cm]{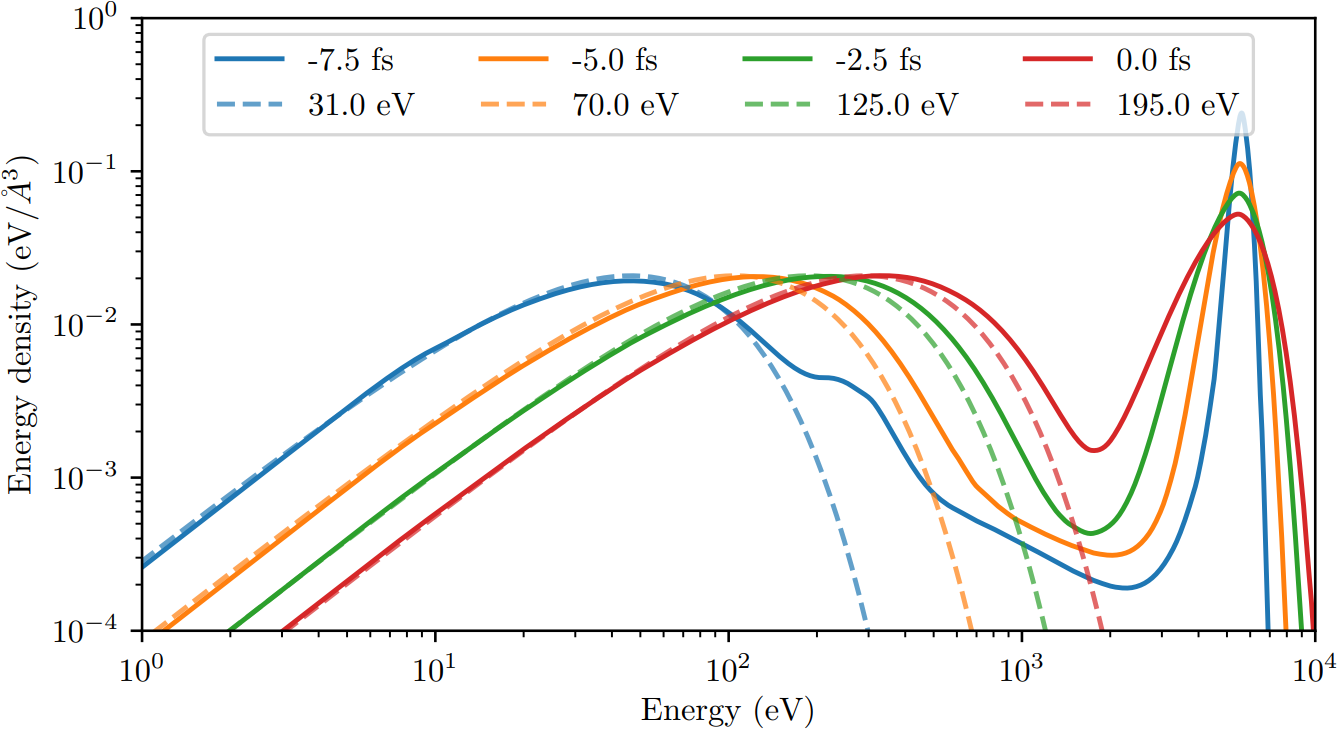}
\caption{Snapshots of the normalized electron energy density distribution in amorphous carbon. Solid lines were predicted by AC4DC, while the dashed lines are the Maxwellian fits by Ref.~\cite{hau-riegeNonequilibriumElectronDynamics2013} based on \texttt{ddcMD}.}
\label{fig:HR}
\end{figure}

Ref.~\cite{hau-riegeNonequilibriumElectronDynamics2013} used the MD code \texttt{ddcMD} to simulate the ionization of amorphous carbon of density 1 g cm$^{-3}$ under a 2 mJ, $6.0$ keV, $10$ fs square pulse focused to a 100 nm FWHM focal size.Modeling with \text{AC4DC} was performedf taking a homogeneous intensity equal to the average in the 100 nm focus.  This corresponds to a photon fluence of  $2.6\times10^{14}$ ph $\cdot$\si{\micro\metre}$^{-2}$, an order of magnitude above the maximum fluence considered elsewhere in this work. \texttt{AC4DC} saw excellent agreement with \texttt{ddcMD} in both the evolution of the carbon ionic states and the free-electron energy distribution. Fig.~\ref{fig:HR} shows the Maxwellian fits to the thermal electrons at several snapshots in time predicted by \texttt{ddcmD}, overlaid on the corresponding normalized electron energy density distributions predicted by \texttt{AC4DC}. %

\subsection{Aluminum sheet -- \texttt{PICLS}}

A system corresponding to an infinite plane \SI{1}{\micro\meter} thick aluminum foil sheet was simulated by Ref.~\cite{royleKineticModelingXray2017} using the particle-in-cell code \texttt{PICLS} under varying conditions. The code tracks individual electrons in space, and determines atomic transitions and impact ionization using Monte Carlo modeling. TBR is incorporated through enforcing an equation of state, which, as highlighted by Ref.~\cite{royleKineticModelingXray2017}, is less accurate than modeling TBR cross-sections. Unlike \texttt{AC4DC}, \texttt{PICLS} models ionization potential depression (IPD). IPD is a stronger effect in denser matter, and so would be expected to play a greater role in solid Al than in biological matter. As the EII rate is increased by IPD, it was expected that \texttt{AC4DC} would underestimate the rate of secondary ionization.  %

We consider two scenarios from this study corresponding to quite different pulse parameters. First, a pulse of 1.70 keV photons with a 80 fs FWHM Gaussian profile and peak intensity of $1.36\times10^{17}$ W cm$^{-2}$. Second, a pulse of 10.0 keV photons with a 20 fs FWHM Gaussian profile and peak intensity of $1\times10^{19}$ W cm$^{-2}$. Note that for the first pulse, the photoelectrons are ejected with a very low energy ($<150$~eV), so the cascades seeded by the 1.4 keV Auger electrons dominate the secondary ionization dynamics.

\onecolumngrid

\begin{figure}[b]

\makebox[\linewidth][c]{
\includegraphics[]{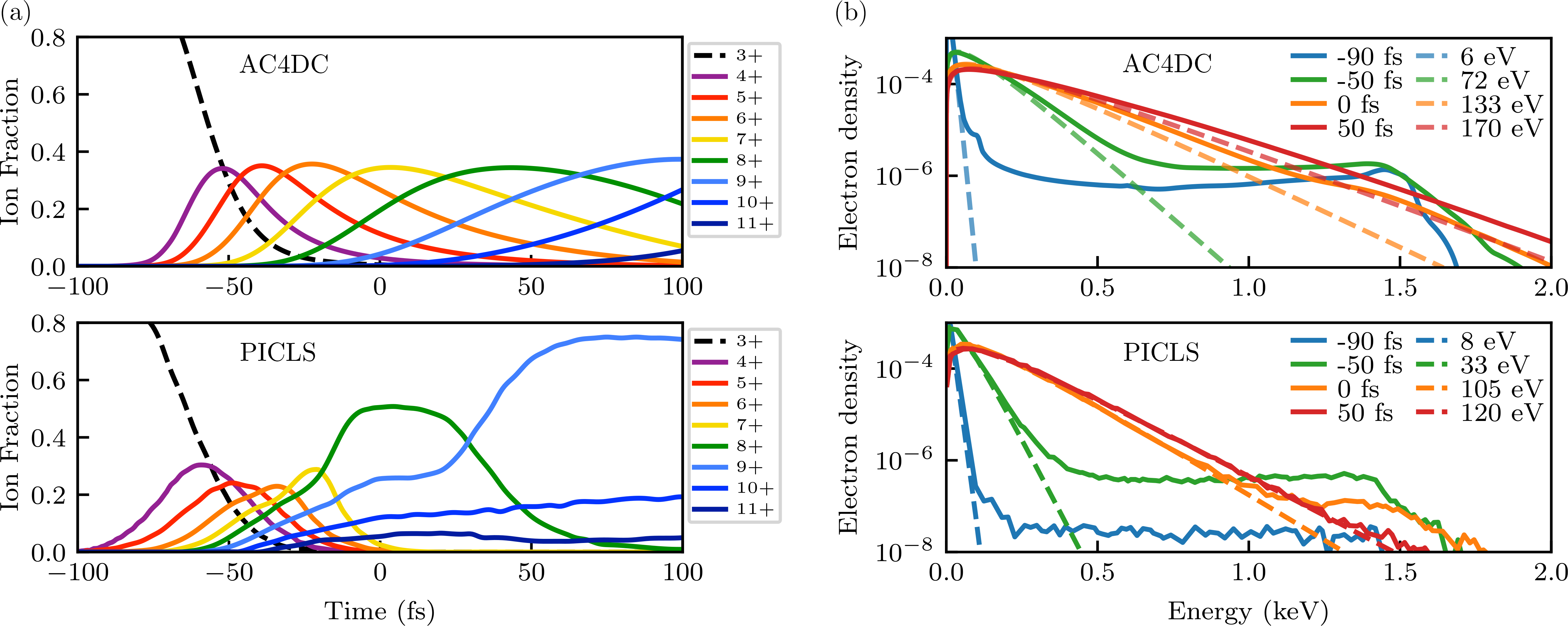}
}
\caption{Predictions by \texttt{AC4DC} (top) and \texttt{PICLS}~\cite{royleKineticModelingXray2017} (bottom) for the (a) ion fractions and (b) normalized electron density of an XFEL-driven solid Al plasma. Maxwellian fits to low-energy electrons are shown as dashed lines.  Both simulations model peak intensities of $1.36\times10^{17}$ W cm$^{-2}$. In the \texttt{PICLS} simulations, the spatial pulse profile is approximately homogeneous.}
\label{fig:royle_b}
\end{figure}

\clearpage 
\twocolumngrid

For the 1.7 keV pulse, the predictions for the evolution of the ion fractions by \texttt{AC4DC} and \texttt{PICLS}, shown in Fig.~\ref{fig:royle_b}(a), see relatively good agreement, similar to that observed in the considered study with another collisional-radiative code, \texttt{SCFLY}. In comparison to PICLS, the ionization rate predicted by \texttt{AC4DC} is overall slightly lower. Inspection of the electron density distributions shown in Fig.~\ref{fig:royle_b}(b) indicates that the electrons thermalize more slowly in the \texttt{AC4DC} simulation. This is consistent with a higher rate of collisional-ionization in the \texttt{PICLS} model. The different level of thermalization between the models limits the physical meaning of a comparison between temperatures. Similar can be said of the electron density distributions for the 10 keV pulse.

\begin{figure}[H]
\includegraphics[]{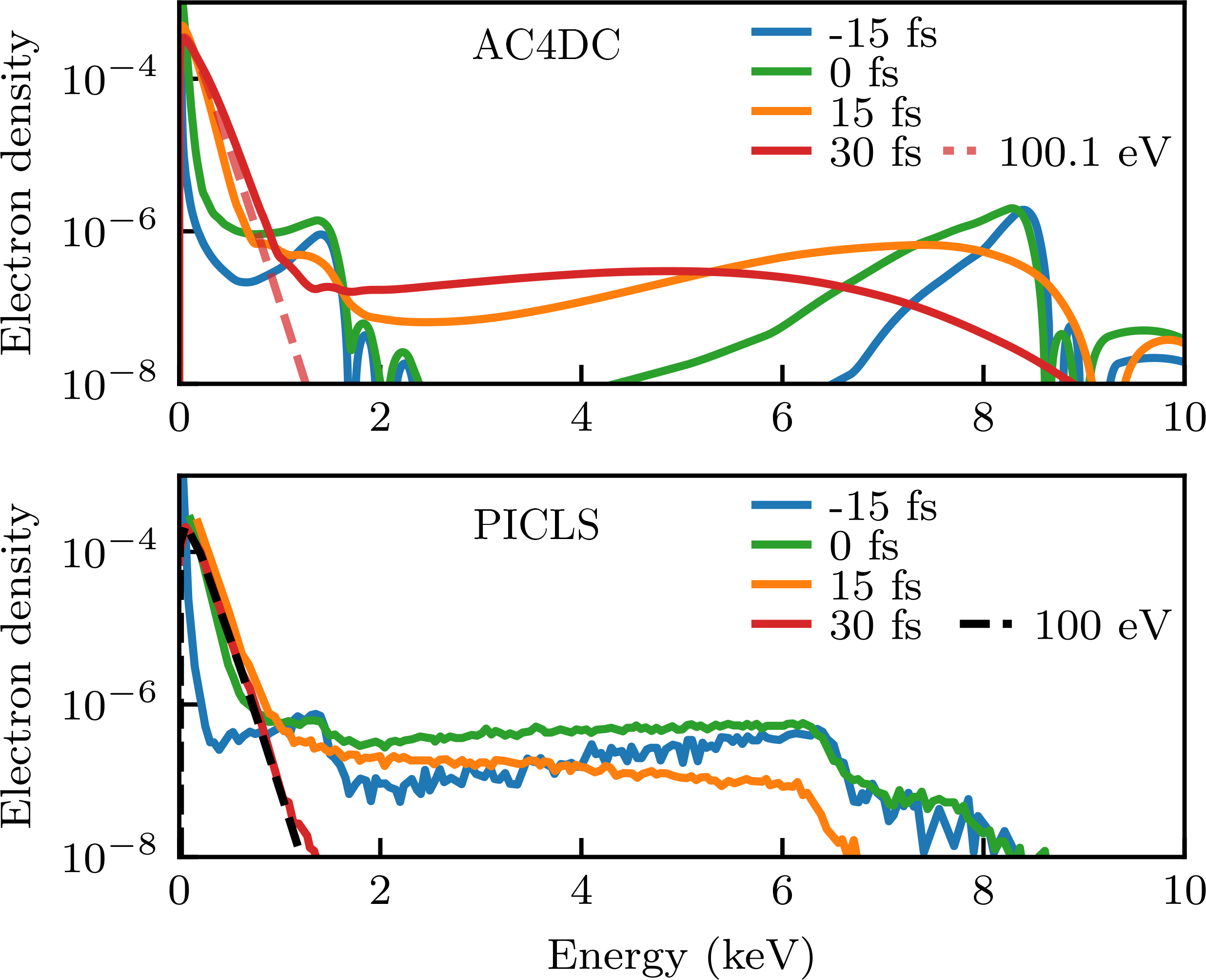}
\caption{Normalized electron densities of Al plasma from \texttt{AC4DC} (top) and \texttt{PICLS} (bottom)~\cite{royleKineticModelingXray2017} simulations under a 10 keV 20 fs FWHM Gaussian pulse. Times are denoted relative to the pulse peak. Both simulations model peak intensities of $1\times10^{19}$ W cm$^{-2}$. Here, \texttt{PICLS} models the radial intensity profile with a super-Gaussian approximately equivalent to a square profile. Dashed lines show the Maxwellian fits to the low-energy electrons at $t=30$ fs ($<2$~keV for the $\texttt{AC4DC}$ trace). }
\label{fig:royle_c}
\end{figure}

For the 10 keV pulse, the models again make similar predictions for the evolution of the free-electron energy distribution, as can be observed in Fig.~\ref{fig:royle_c}. There is a surprisingly good match to the temperature of the thermal electrons at $t=+30$ fs. However, this is again a limited comparison as the low-energy electrons have not completely relaxed to a Maxwellian in the \texttt{AC4DC} simulation, while they are nearly completely thermalized in the \texttt{PICLS} model. The low-energy electrons remained unrelaxed at $t=+100$ fs in the \texttt{AC4DC} simulation (with a best-fit Maxwellian of 213.5 eV).

\newpage 
\subsection{Glycine crystal -- \texttt{XMDYN}}

\begin{figure}
\includegraphics[]{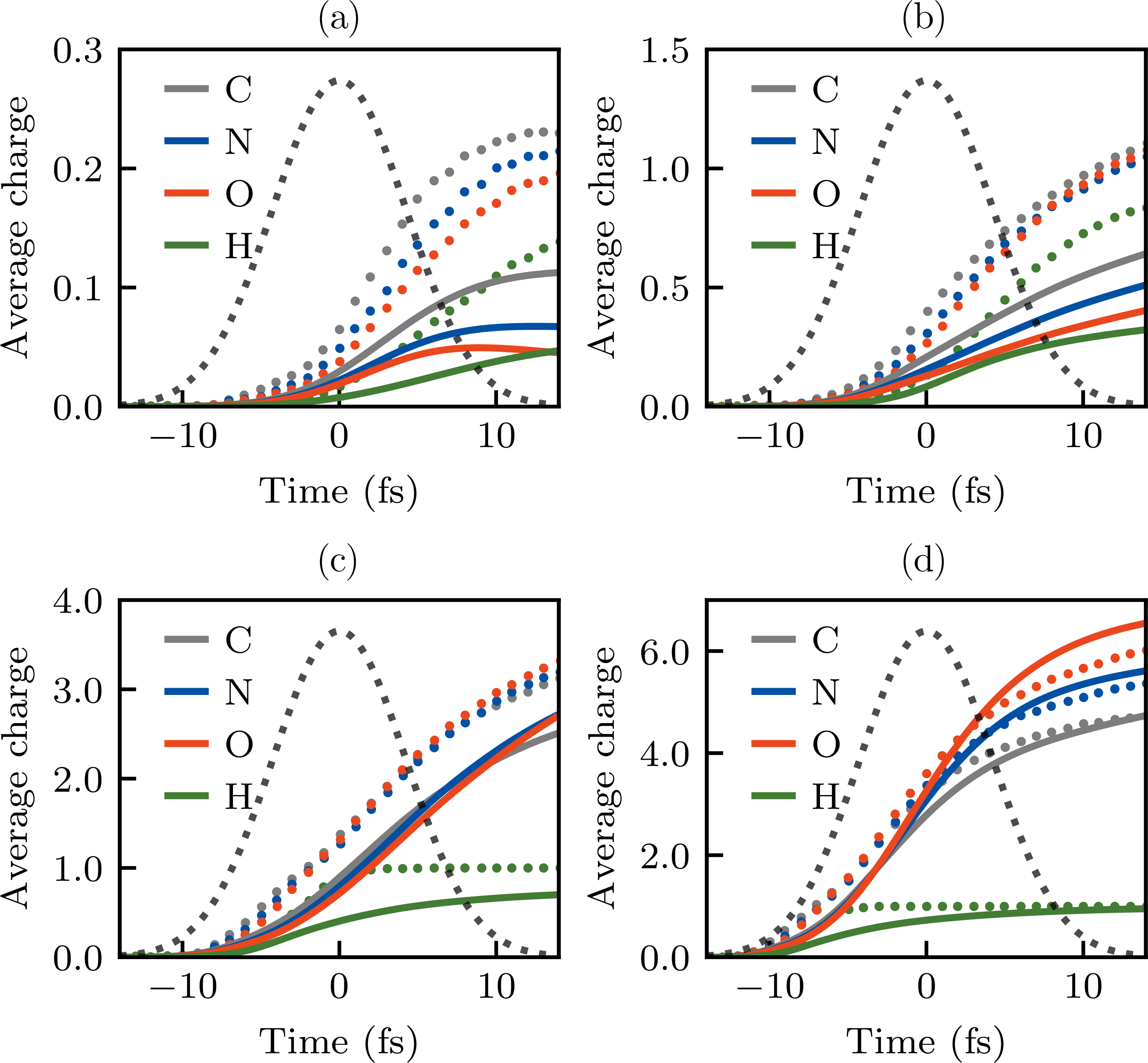}[H]

\caption{Predictions by \texttt{XMDYN}~\cite{abdullah_MD_non-equilibrium2018} (dots) and  \texttt{AC4DC} (lines) for the average charges in the glycine crystal under different peak intensities: (a) $1.5\times10^{18}$ W cm$^{-2}$, (b) $1.5\times10^{19}$ W cm$^{-2}$, (c) $1.5\times10^{20}$ W cm$^{-2}$, (d) $1.5\times10^{21}$ W cm$^{-2}$.}
\label{fig:glycine}
\end{figure}

Ref.~\cite{abdullah_MD_non-equilibrium2018} used the Monte Carlo MD code \texttt{XMDYN} to model radiation damage in a glycine (H$_{5}$C$_{2}$NO$_{2}$) crystal. The study modeled pulses of 10 keV photons with 10 fs FWHM Gaussian temporal profiles, at four different peak intensities, between $1.5\times10^{18}$--$1.5\times10^{21}$ W cm$^{-2}$. The simulation volume contained 1050 atoms, under periodic boundary conditions. The results of these simulations are shown in  Fig.~\ref{fig:glycine}. The evolution of charges by element in equivalent \texttt{AC4DC} simulations, also shown in Fig.~\ref{fig:glycine}, was in good agreement with this modeling for the two higher fluence cases. In the two lower fluence cases, \texttt{XMDYN} predicts a higher level of ionization. %

\subsection{Silicon crystal -- non-Maxwellian plasma code}

\begin{figure}
\includegraphics[width=8cm]{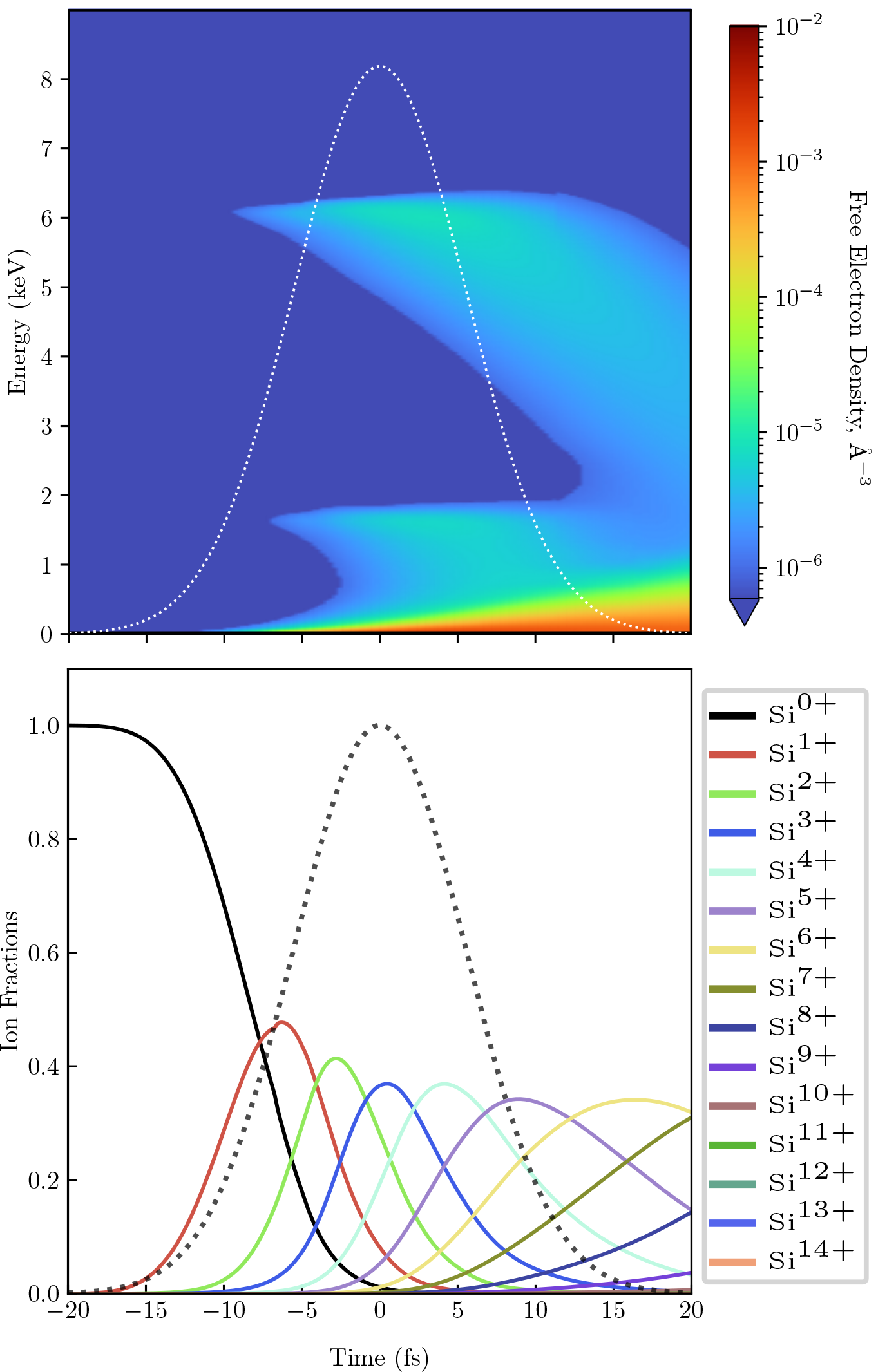}
\caption{Electron energy density (top) and ion fractions (bottom) in Si under an 8 keV 13 fs FWHM Gaussian pulse of fluence $1.6\times10^{5}$ J$\cdot$cm$^{-2}$. The dotted lines show the pulse profile.}
\label{fig:silicon}
\end{figure}

We modeled the conditions a silicon crystal was subjected to in simulations performed by Ref.~\cite{leonovTimeDependenceXray2014}. The considered study used an atomic code that was shown to provide good agreement with the HF method used by AC4DC. The physical model for the plasma dynamics is very similar to \texttt{AC4DC}, though the numerical approaches differ significantly. \texttt{AC4DC} was used to model the 4 keV and 8 keV pulses considered in this study, corresponding to energies near and from the 1.8 keV Si K-edge. \texttt{AC4DC} predicted a much higher rate of ionization and differing electron density continuums. Results for the 8 keV pulse are shown in Fig.~\ref{fig:silicon}.

\subsection{Lysozyme.Gd -- RADDOSE-XFEL}
\label{app:comparisonsRADDOSE}

\texttt{RADDOSE-XFEL}~\cite{dickersonRADDOSE-XFEL2020} is a module of \texttt{RADDOSE-3D}~\cite{BuryRADDOSE3D2017} designed to measure the dose absorbed in XFEL experiments by tracking collisional ionization cascades. Unlike \texttt{AC4DC}, the code does not model high-density, strong-ionization effects such as TBR and the depletion of atomic orbitals. Since primary ionization drives the damage-seeding effect described in this study, the code was expected to reflect an outsized impact by heavy elements on the absorbed dose.

\newpage 

Table~\ref{tab:RADDOSE} summarizes the damage predictions of RADDOSE-XFEL and \texttt{RADDOSE-3D} applied to targets equivalent to the lysozyme.Gd target and its toy variants described in Sec.~\ref{sec:lysozymeGd}. Values are scaled based off a sample of $2\times10^7$ individually simulated photons, as described in Ref.~\cite{dickersonRADDOSE-XFEL2020}. The doses predicted by \texttt{RADDOSE-3D} are consistently higher than predicted by \texttt{RADDOSE-XFEL} because \texttt{RADDOSE-3D} assumes the energy of photons are instantaneously absorbed after photoionization~\cite{dickersonRADDOSE-XFEL2020}. The influence of heavy atoms is underestimated by this approximation; adding sulfur (light control to lysyozyme, water) and salt (lysozyme.Gd, 0.1 M Gd to lysozyme.Gd, 0.1 M Gd 10\% NaCl) magnifies the dose by a much larger factor when the more accurate \texttt{RADDOSE-XFEL} model is used.  Notably, accounting for sulfur increases the absorbed dose and average number of non-H ionizations by 49\% and 45\%, respectively. This effect is stronger than predicted by the \texttt{AC4DC} simulations, likely because \texttt{RADDOSE-XFEL} does not model the ionic states and so does not capture the suppression of EII that occurs as the target becomes increasingly ionized.

\texttt{RADDOSE-XFEL} predicts that accounting for Gd (lysozyme, water to lysozyme.Gd, 0.1 M Gd) does not increase ionization, but increases the dose by an amount comparable to that predicted by \texttt{RADDOSE-3D}. This result should be taken as unphysical.  \texttt{RADDOSE-XFEL} models instantaneous energy deposition similar to \texttt{RADDOSE-3D} for many ionization processes of Gd. Indeed, Fig.~\ref{fig:cascade_electrons_by_element} indicates the contribution of cascades seeded by Gd to global ionization should be similar to those seeded by Cl in the 10\% NaCl solvent.

\onecolumngrid

\begin{table}[H]
  \centering
  \begin{threeparttable}
    \caption{Predictions of \texttt{RADDOSE-XFEL} and \texttt{RADDOSE-3D}. The target column gives the solvent. Note the right column shows the average number of ionizations for non-H atoms only. Pulse parameters and targets are equivalent to those given in Sec.~\ref{subsec:ionization_modes}.}
    \label{tab:RADDOSE}
    \begin{tabular}{llccc}
      \toprule
      Protein & Solvent (35.1\% v/v) & \begin{tabular}{@{}c@{}}\texttt{RADDOSE-3D} \\  dose (MGy)\end{tabular} &  \begin{tabular}{@{}c@{}}\texttt{RADDOSE-XFEL} \\  dose (MGy)\end{tabular} & \begin{tabular}{@{}c@{}}\texttt{RADDOSE-XFEL} \\  avg. non-H ionizations\end{tabular}  \\
      \midrule
      Light-atom control, & water &  1931.4  & 325.7  & 1.278 \\
      Lysozyme, & water & 2338.8  & 483.6  & 1.856 \\
      Lysozyme.Gd, & 0.1 M Gd & 2923.2  & 1088.6\tnote{*}  & 1.859\tnote{*} \\
      Lysozyme.Gd, & 0.1 M Gd 10\% NaCl  & 3476.6  & 1304.8\tnote{*}  & 2.779\tnote{*} \\
      \bottomrule
    \end{tabular}
    \begin{tablenotes}
      \item[*] \texttt{RADDOSE-XFEL} assumes instantaneous dose absorption for some Gd ionization processes.
    \end{tablenotes}
  \end{threeparttable}
\end{table}

\twocolumngrid 

\clearpage
\onecolumngrid

\section{Supplementary Figures}
\makeatletter
\renewcommand \thesection{S\@arabic\c@section}
\renewcommand\thetable{S\@arabic\c@table}
\renewcommand \thefigure{S\@arabic\c@figure}
\makeatother
\setcounter{figure}{0}    
\setcounter{table}{0}   

\twocolumngrid

\begin{figure}[H]
\includegraphics{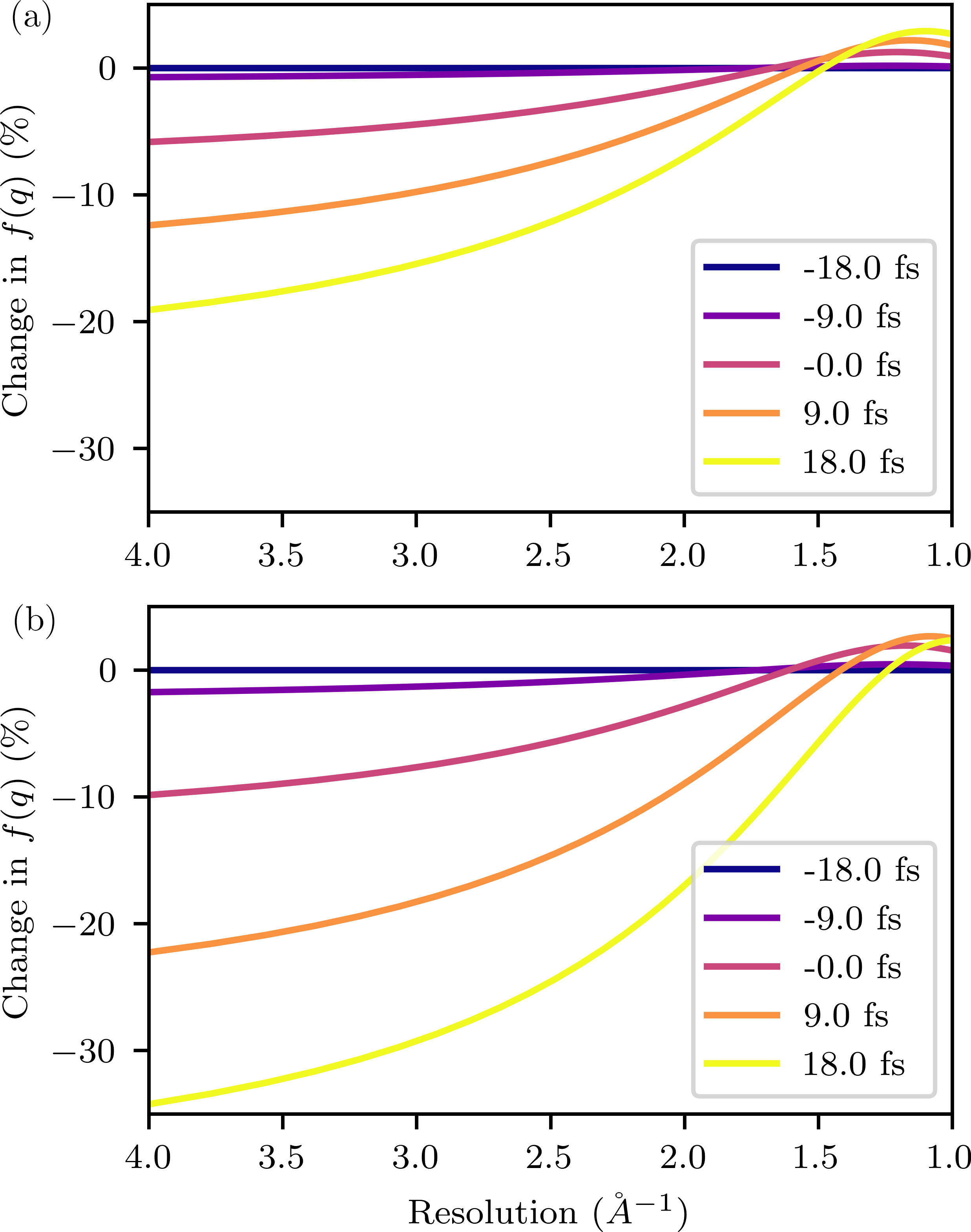} \label{fig:atomic_form_factor}
\caption{Evolution of the scattering amplitude of carbon atoms in the lysozyme.Gd protein, under the illumination conditions of Sec.~\ref{subsec:ionization_modes}. (a) Simulation where only light-atom ionization is modeled. (b) Simulation where all ionization is modeled. Traces correspond to the `average' carbon atom in the protein at the denoted times, defined as a carbon atom with the average orbital occupancies of all carbon atoms. The atomic form factor $f(q)$, where $q$ is the momentum transfer, is defined as the Fourier transform of the atom's electron density. The vertical axis gives the change in $f(q)$ for the average carbon atom relative to the form factor of a carbon atom in its neutral ground state. The horizontal axis gives the resolution ($2\pi/q$) that corresponds to the scattering angle.} 
\end{figure}

\begin{figure}[H]
\includegraphics[width=3.49751in]{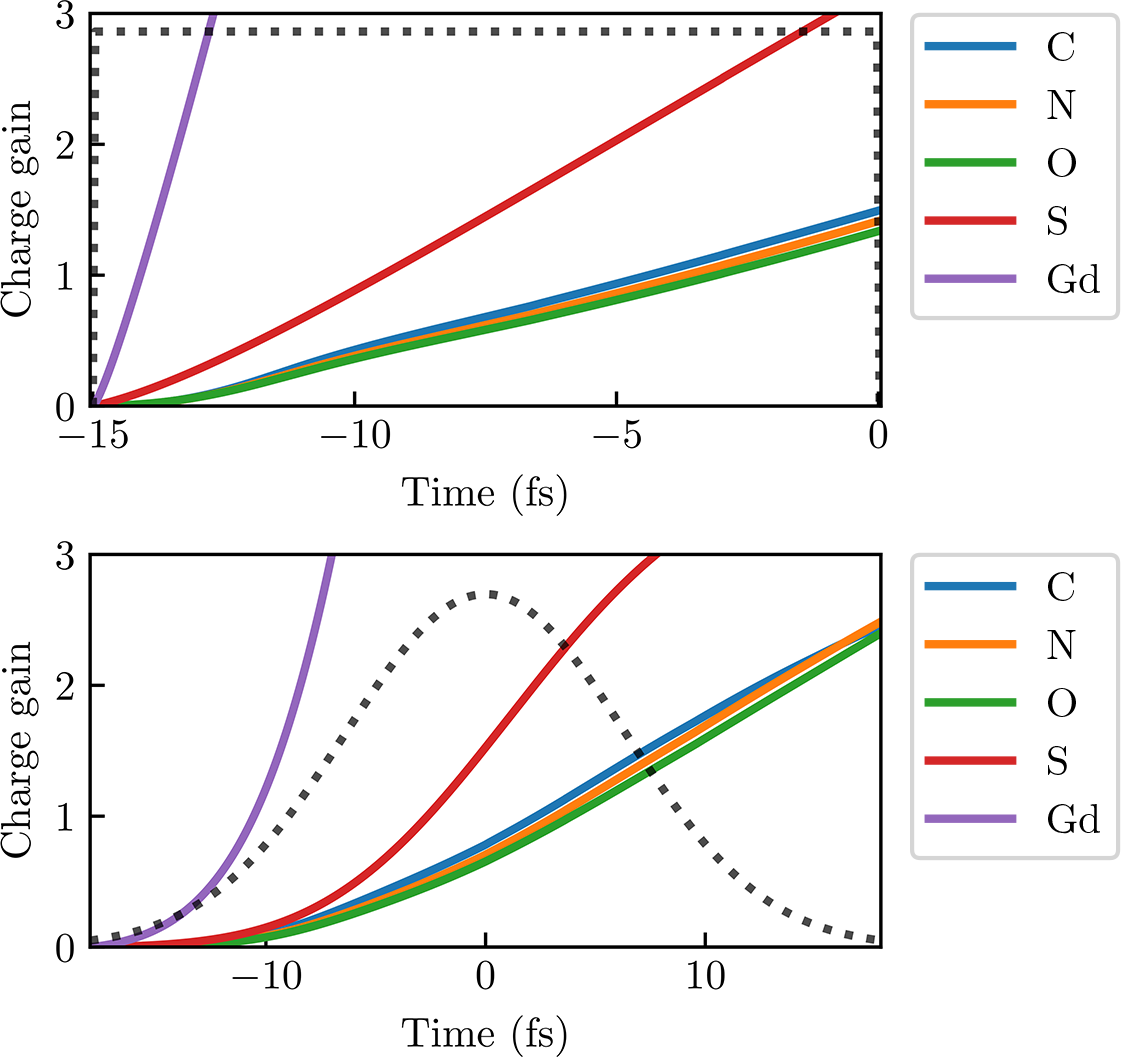}
\caption{Effect of the choice of pulse profile used in simulating the dynamics for the lysozyme.Gd crystal with 10\% NaCl solvent considered in Sec.~\ref{sec:lysozymeGd}. Plots show the evolution of the average charges for elements present in the protein under the Gaussian and square pulse profile idealizations for a 15 fs FWHM pulse, as represented by the dotted lines. Both pulses have a fluence of $1.75 \times 10^{12}$ 7.112 keV ph$\cdot$\si{\micro\metre}$^{-2}$. By $t=0$, the target is in a more ionized state under the Gaussian pulse.} %
\label{fig:pulse_profile}
\end{figure}

\begin{figure}[H]
	\centering
        \includegraphics[]{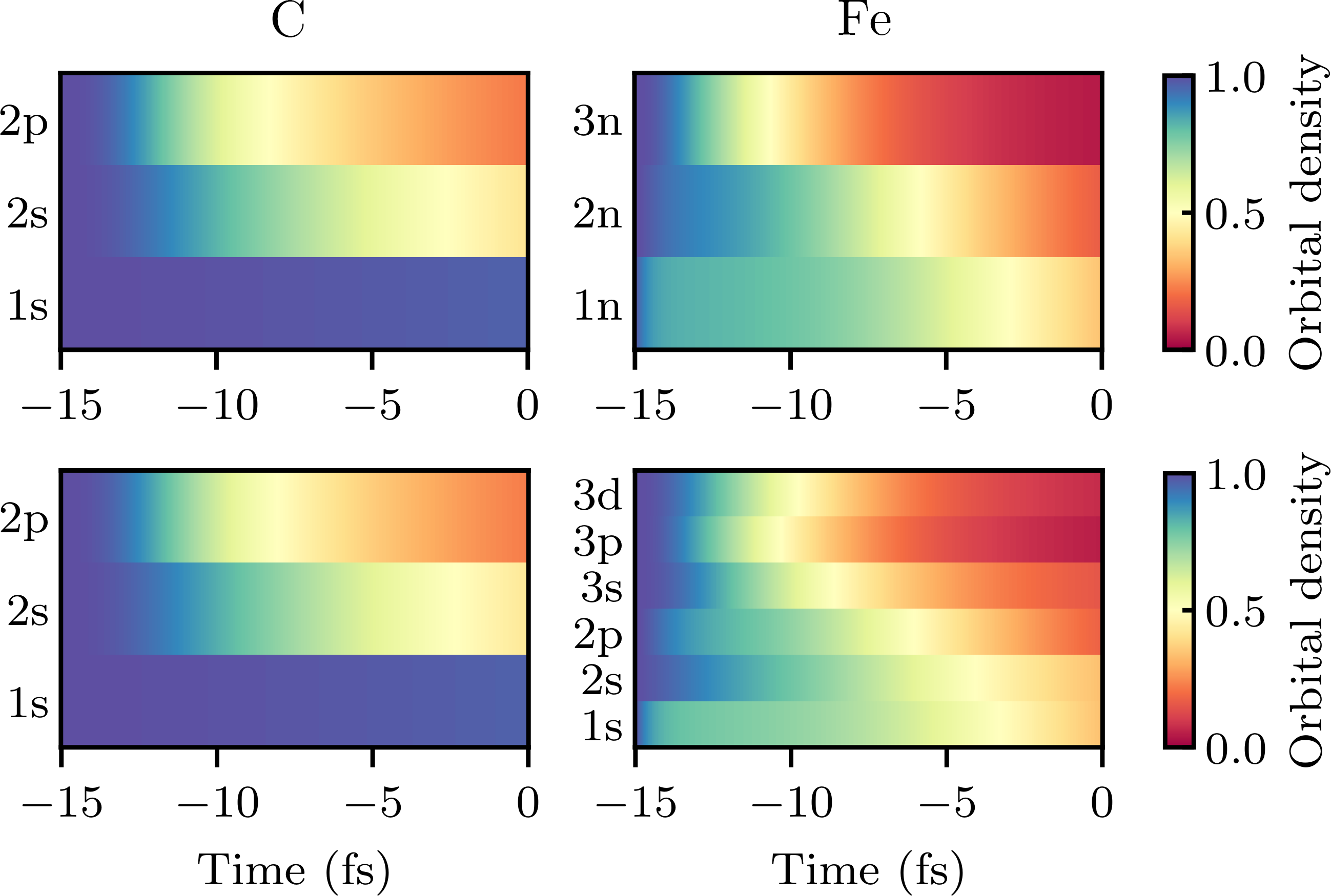}       
	\caption{Occupancy of C and Fe within the Fe-doped protein, with (top) and without (bottom) the single-shell approximation that was applied in this study to atoms heavier than Fe. The pulse was modeled with a 15 fs square temporal profile, and a fluence of $10^{13}$ 10 keV ph$\cdot$\si{\micro\metre}$^{-2}$. TBR was disabled in these simulations. The single-shell approximation has a negligible impact on the ionization of C. 
        }
    \label{fig:fe_single_shell_comparison}
\end{figure}

\clearpage

\onecolumngrid

\begin{figure}[H]
	\centering
        \includegraphics[width=15cm]{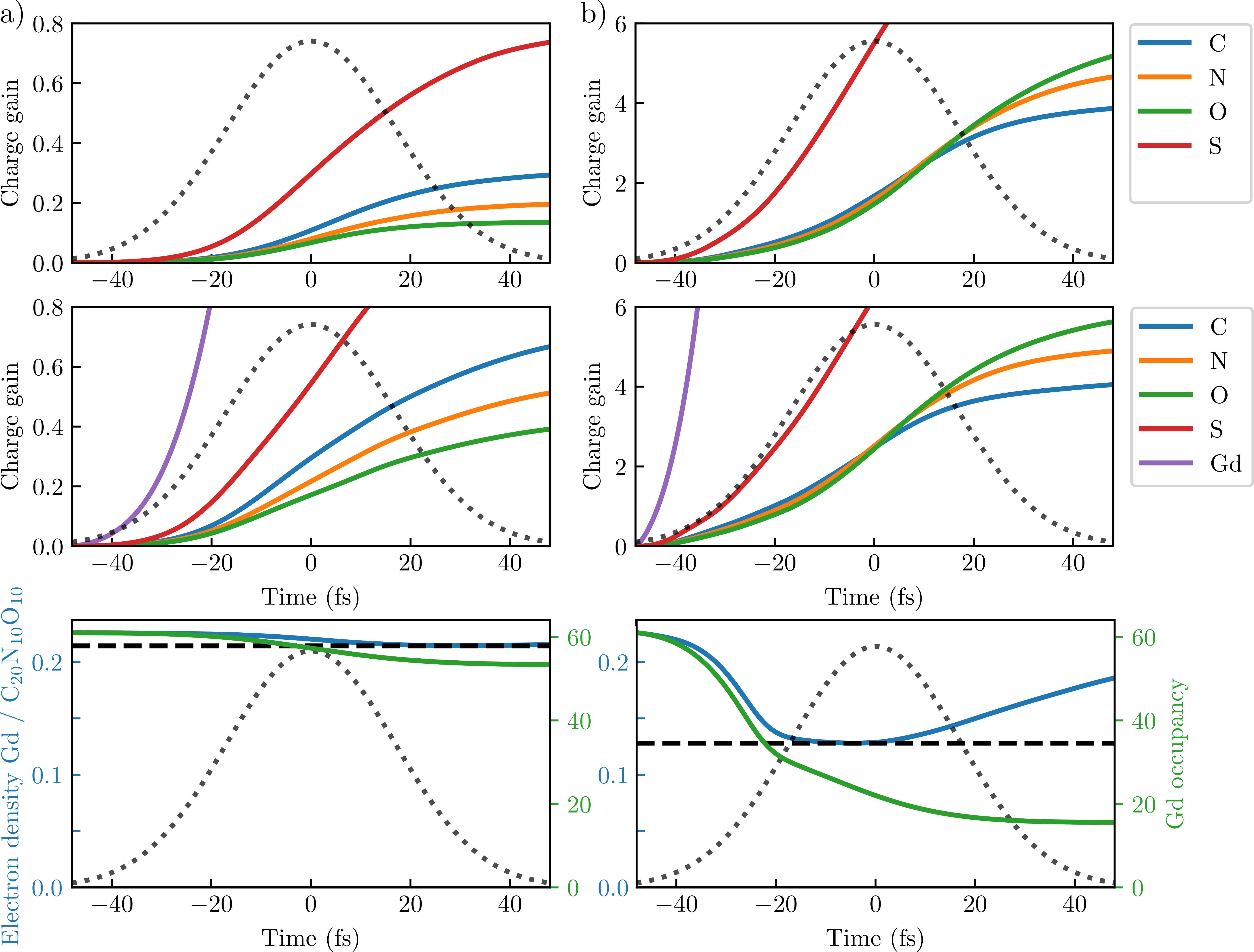}
        
\caption{Effect of Gd ions on the ionization of lysozyme.Gd in the (a) low and (b) high fluence experiments performed by Ref.~\cite{galli}. Pulses were simulated with 40 fs FWHM Gaussian profiles, and (a) $1.3 \times 10^{11}$ 8.48 keV ph$\cdot$\si{\micro\metre}$^{-2}$ or (b) $7.8 \times 10^{12}$ 8.48 keV ph$\cdot$\si{\micro\metre}$^{-2}$ fluence. The photon energy is above the L-edge of Gd (modeled as 7.42 keV). Plots in the top row correspond to lysozyme in water as solvent, and plots in the middle row correspond to the true composition: lysozyme.Gd in 0.1 M Gd 8\% NaCl solvent. The presence of Na$^+$, Cl$^-$, and Gd$^{3+}$ ions (middle row) increases the ionization of the light atoms. Plots in the bottom row show the evolution of the Gd occupancy and EDR. The horizontal dashed line shows the EDR and Gd occupancy based on the ionization theoretically predicted for each element in the original study, where the quantities are assumed to be commensurate. Note that Gd occupancy is proportional to EDR neglecting light atom damage. The EDR and Gd occupancy diverge significantly during the pulse due to the substantial light-atom ionization.}
    \label{fig:galli_charge}
\end{figure}

\begin{figure}[H]
\makebox[\linewidth][c]{
    \includegraphics[]{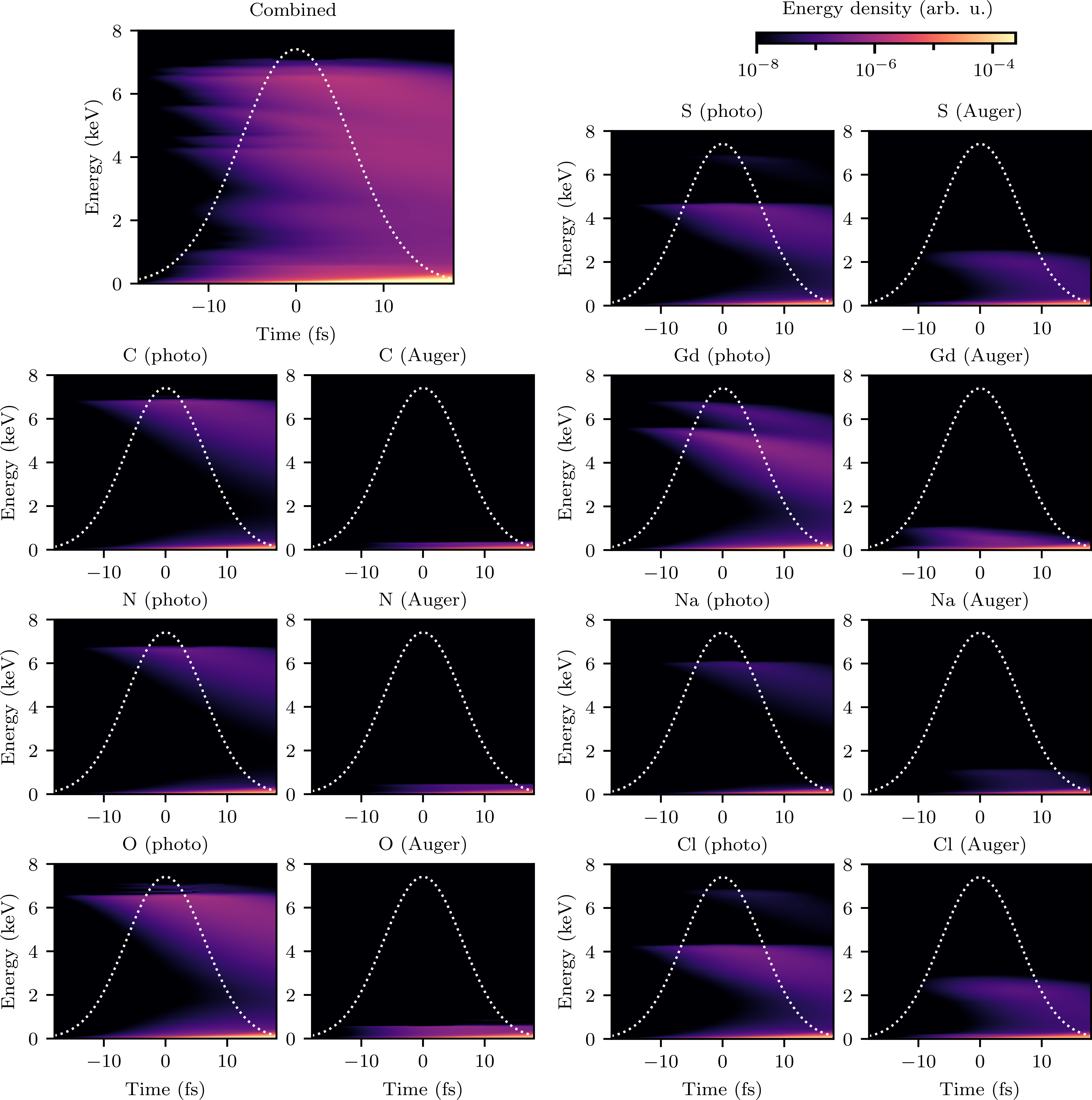}
}
\caption{Contributions of each element to the free-electron cascades in the lysozyme.Gd crystal under the illumination conditions of Sec.~\ref{subsec:ionization_modes}. Each plot shows the energy density of electrons freed from EII cascades seeded by Auger electrons or photoelectrons from an element, as denoted above the plot. Secondary ionization of heavy elements is ignored. The continuums include the contribution of the primary electrons. The `combined' plot corresponds to the complete free-electron continuum.}
\label{fig:split_continuum_lys_Gd_solvated}
\end{figure}

\twocolumngrid

\end{document}